\documentclass[aps,prb,superscriptaddress,twocolumn,floatfix,notitlepage]{revtex4-1}
\usepackage[percent]{overpic}
\usepackage{graphicx,graphics}
\usepackage{dcolumn}
\usepackage{amsmath,amssymb,amsfonts}
\usepackage{wasysym}
\usepackage{latexsym,verbatim}
\usepackage{bm}
\usepackage{color}
\usepackage{ulem}
\usepackage[framemethod=default]{mdframed}
\usepackage[breaklinks=true,colorlinks,citecolor=blue,linkcolor=blue,urlcolor=blue]{hyperref}
\usepackage[makeroom]{cancel}
\usepackage{fancybox}
\usepackage{bbm}
\usepackage{physics}
\usepackage{tikz}
\usepackage{bbold}
\usepackage[inline]{enumitem}

\begin{document}

\author{Thomas Benjamin Smith}
\email{tommy.smith023@gmail.com}
\address{Department of Physics and Astronomy, School of Natural Sciences, Faculty of Science and Engineering, University of Manchester, Oxford Road, Manchester, M13 9PY, United Kingdom.}
\author{Alessandro Principi}
\address{Department of Physics and Astronomy, School of Natural Sciences, Faculty of Science and Engineering, University of Manchester, Oxford Road, Manchester, M13 9PY, United Kingdom.}

\title{Emergent Non-Hermitian Edge Polarisation in an Hermitian Tight-binding Model}

\begin{abstract}
We study a bipartite Kronig-Penney model with negative Dirac-delta potentials that may be used, amongst other models, to interpret plasmon propagation in nanoparticle arrays. Such a system can be mapped into a Su-Schrieffer-Heeger-like model however, in general, the overlap between `atomic' wavefunctions of neighbouring sites is not negligible. In such a case, the edge states of the finite system, which retain their topological protection, appear to be either attenuated or amplified. This phenomenon, called ``edge polarisation'', is usually associated with an underlying non-Hermitian topology. 
By investigating the bulk system, we show that the resulting tight-binding eigenvalue problem may be made to appear non-Hermitian in this physical `atomic' (lattice-site) basis. The resulting {\it effective} bulk Hamiltonian possesses ${\cal PT}$-symmetry and its topological invariant, interpreted in terms of a non-Hermitian classification, is found to be given by a bulk winding number of $\mathbb{Z}$-type.
The observation of edge polarisation, through the established bulk-boundary correspondence, is then interpreted as an emerging non-Hermitian skin-effect of the {\it effective} bulk Hamiltonian.
Therefore, the overlap matrix generates non-Hermitian-like effects in an otherwise Hermitian problem; a general fact applicable to a broader range of systems than just the one studied here. 
\end{abstract}


\maketitle

\section{Introduction}

The field of non-Hermitian topology continues to grow at pace\cite{Kunst:2018,Yao:2018,Liu:2019,Lee:2019,Lopez:2019,Gong:2018,Zhao:2019,
Yuce:2015,Yuce1:2018,Yuce2:2018,Borgnia:2020,Turker:2018,Ghatak:2020}. In general, the study of non-Hermitian quantum systems has historically been eschewed on the basis of non-physicality\cite{Bender:2007,Dirac:1947,Schrodinger:1926}. Indeed, in paradigmatical quantum mechanical Hamiltonian eigenvalue problems, the enforced Hermiticity: \begin{enumerate*}[label=(\roman*)] \item guarantees the reality of the resultant energy eigenvalues, \item imposes the required unitarity of the time-evolution of the system, and \item ensures that left and right eigenvectors are identically equivalent.\end{enumerate*}

On the other hand, the energy eigenvalues of a non-Hermitian Hamiltonian may in fact be, in general, complex-valued thereby violating unitarity. Furthermore, the equivalency between left and right eigenvectors is also broken in such a case.

As the ideas of quantum mechanics permeate into more complex systems, the fixation upon the Hermiticity of eigenvalue problems has been relaxed as the constituent excitations are capable of temporal decay, in the cases of, e.g. excitons\cite{Gao:2015}, plasmons\cite{Varguet:2019,Cortes:2020}, and phonons\cite{Lu:2017,Wang2:2018}, and/or variations in their phase through gain and loss in the cases of, e.g. photons\cite{Bender:2016,Makris:2008} and plasmons\cite{Ke:2017}.

As a result, the field of non-Hermitian topological protection has been delved into in earnest with a more general 36-fold way of non-Hermitian topological invariants being developed\cite{Gong:2018,Kawabata1:2019}.

In such systems, radical departures from the basic Hermitian model are observed. For example: \begin{enumerate*}[label=(\roman*)] \item the emergence of exceptional points whereat the bulk bands develop imaginary components\cite{Kato:1976,Foa_Torres:2019,Kawabata2:2019,Yoshida:2019}, \item a richer variety of topological protections due to the increased number of possible symmetries\cite{Kawabata1:2019,Kawabata3:2019}, and \item the non-Hermitian skin effect (both normal and anomalous)\cite{Ghatak:2019,Yuce:2019,Longhi:2019} within which the phenomenon of edge polarisation\cite{Ghatak:2019} appears. \end{enumerate*} The latter consists in the attenuation and amplification of {\it topologically protected} and {\it degenerate} edge states of one-dimensional bipartite chains by on-site balanced gain and losses. Such behaviour is especially appealing because it could be used to experimentally detect signatures of non-Hermitian topology\cite{Takata:2018,Jin:2017,Wang:2020,Zhan:2017}, for example in the aforementioned physical systems.

In this paper we ask the question: is it possible to observe any effect as those described in the immediate above in perfectly Hermitian systems? We answer affirmatively by showing that, when the (usually neglected) overlap matrix between neighbouring localised orbitals is taken into account, the edge states of a conventional Su-Schrieffer-Heeger (SSH) model~\cite{SSH:1979,SSH:1980} can appear attenuated and amplified while retaining their topological protection.

The attenuation and amplification of degenerate edge states in one-dimensional {\it Hermitian} lattices, {\it i.e.} the fact that their wavefunctions have unequal weights on the outer sites of the chain, is conventionally attributed to the lack of chiral symmetry, which also splits their energetic degeneracy\cite{Asboth:2016}. This is the case in (e.g.) the celebrated SSH model. Only in the absence of an on-site potential difference between the two atomic sites within the unit cell will the edge states appear degenerate at zero (mid-gap) energy and be equally shared by the outer sites of the chain.

In this case, the presence or absence of such states can be deduced through the concept of bulk-boundary correspondence by computing a bulk topological invariant. In the presence of chiral symmetry, this resides within the winding number of the off-diagonal Hamiltonian element~\cite{Asboth:2016} or (equivalently) the {\it Hermitian} Zak phase~\cite{Ghatak:2019,Zhang:2019,Zak:1989}:
\begin{equation}\label{eq:herm_zak}
\theta_{\cal Z}=i\int_{-\pi/d}^{+\pi/d}dk \langle\psi_k|
\partial_k\psi_k\rangle
~.
\end{equation}
Here $|\psi_k\rangle$ is the periodic part of the Bloch wavefunction while $d$ is the length of the unit cell. In the presence of chiral symmetry, the winding number (or equivalently the Zak phase) is quantised~\cite{Asboth:2016,Zak:1989}. The system is therefore characterised by this $\mathbb{Z}$ topological invariant, which is equivalent to the number of edge modes through the bulk-boundary correspondence. An on-site chiral-symmetry-breaking potential leads to the destruction of the topological protection and perfect degeneracy, the localisation of the edge states to either boundary site, and to the simultaneous loss of a well-defined bulk invariant. 

In the non-Hermitian analogue of the SSH model with on-site balanced gains and losses~\cite{Jin:2017,Yuce2:2018}, edge states retain their degeneracy and topological protection, while appearing at the same time to have different weights on the outermost sites. Since the Hamiltonian in this case belongs to the BDI class, according to the periodic table of invariants~\cite{Gong:2018} the system is characterised by a $\mathbb{Z}\oplus\mathbb{Z}$ invariant. The topological invariant is there composed of two numbers~\cite{Yuce:2019,Song:2019}, one of which is equivalent to the Zak phase (modulo $\pi$), while the other is the winding number of the imaginary part of the energy (modulo $\pi$). 

Here we show that the overlap between wavefunctions belonging to neighbouring sites, normally neglected in topological {\it Hermitian} tight-binding models, leads to a tendency of edge states to localise to either end. However, at odds with the conventional phenomenology described above, we show that the overlap does not break the chiral symmetry, nor lift the degeneracy of the edge states, nor lead to a non-quantisation of the bulk invariant. By forcing the eigenvalue problem to be expressed in the sublattice basis, we also show that it is possible to interpret this fact as an emergent non-Hermitian effect. This further shows that the amplification and attenuation of edge states should be cautiously used as an indicator of non-Hermitian topology in open systems, since it can occur also in perfectly Hermitian problems.

\begin{figure}
	\centering
	\begin{tikzpicture}[scale=0.935, every node/.style={scale=0.935}]
	\draw (-.55,-.5)--(2.525,-.5)--(2.525,.5)--(4.525,.5)--(4.525,-.5)--(7.5,-.5);
	\filldraw (-.5,0)--(-.5,-1.75)--(-.55,-1.75)--(-.55,0)--(-.5,0);
	\filldraw (2.5,0)--(2.5,-1.75)--(2.55,-1.75)--(2.55,0)--(2.5,0);
	\filldraw (4.5,0)--(4.5,-1.75)--(4.55,-1.75)--(4.55,0)--(4.5,0);
	\filldraw (7.5,0)--(7.5,-1.75)--(7.55,-1.75)--(7.55,0)--(7.5,0);
	\draw[dotted] (1,-2.)--(6,-2.)--(6,.75)--(1,.75)--(1,-2.);
	\draw[dotted] (-.5,0)--(7.55,0);
	\draw[<->] (1,-1.9)--(6,-1.9) node at (3.5,-1.9) [anchor=south] {$d$};
	\draw[<->] (2.55,-.25)--(4.5,-.25) node at (3.525,-.25) [anchor=north] {$v$};
	\draw[<->] (4.55,-.875)--(7.5,-.875) node at (6.025,-.875) [anchor=north west] {$w$};
	\draw[->] (-.65,0)--(-.65,-1.75) node at (-.65,-.875) [anchor=east] {$V$};
	\draw node at (1,.75) [anchor=south] {$x_{\rm L}$};
	\draw node at (-.525,.75) [anchor=south] {$x_{B0}$};
	\draw node at (2.525,.75) [anchor=south] {$x_{A1}$};
	\draw node at (4.525,.75) [anchor=south] {$x_{B1}$};
	\draw node at (7.525,.75) [anchor=south] {$x_{A2}$};
	\draw node at (6,.75) [anchor=south] {$x_{\rm R}$};
	\draw[->] (4.25,0)--(4.25,.5) node at (4.25,.25) [anchor=east] {$V_v$};
	\draw[->] (4.75,0)--(4.75,-.5) node at (4.75,-.25) [anchor=west] {$V_w$};
	\draw node at (7.55,0) [anchor=west] {$V_0$};
	\end{tikzpicture}
	\caption{The general bipartite unit-cell under consideration. The Dirac-delta potentials have their strengths defined in relation to $V_0=0$. Then, as long as $V_v$ and $V_w$ are varied symmetrically then the Dirac-delta strength remains constant with respect to $V_0$.}
	\label{fig:unitcell}
\end{figure}
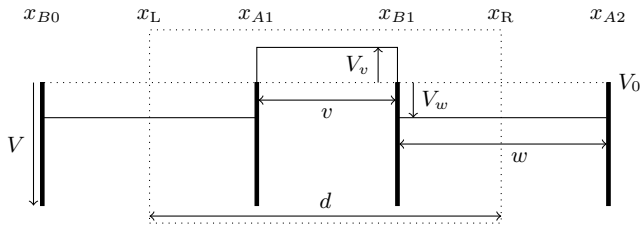

To be concrete, the system we consider herein is a bipartite Kronig-Penney\cite{Kronig:1931,KP:1931} model with negative-strength Dirac-delta potentials, as shown in Fig.~\ref{fig:unitcell}, that is constructed in order to effectively mimic the SSH model\cite{Asboth:2016,Smith:2019}. The bipartition may be achieved by either varying the distances or the baseline potentials between the two Dirac-deltas. It is crucial that the strengths of the Dirac-delta potentials themselves are made to be identical. In this way, no bulk on-site potential is introduced that would trivially distinguish the sublattices within the tight-binding model. As such, chiral symmetry is not destroyed trivially from the get-go. Besides being simple, this model is applicable to a variety of problems. Not least to describe plasmon propagation in both metal gratings~\cite{Smith:2020,Kocabas:2009,Senlik:2009} and nanoparticle arrays\cite{Yousefi:2019,Pocock:2018,Downing:2018}, both of which are prime candidates for the experimental observation of non-Hermitian topology\cite{Ke:2017,Fu:2020,Budich:2020}. Our work therefore makes manifest the danger of using edge polarisation as an experimental diagnostic of non-Hermitian topology, since it may lead to a misinterpretation of results.

The bulk tight-binding time-independent Schr{\"o}dinger equation that applies to the Kronig-Penney model as shown in Fig.~\ref{fig:unitcell} is derived in Apps.~\ref{app:A} and~\ref{app:B}, and reads: $H(k)\psi_k=E(k)S(k)\psi_k$. Explicitly, it takes the following form:
\begin{equation}\label{eqn:TBSE}
\begin{pmatrix}
\epsilon&h(k)
\\
h^*(k)&\epsilon
\end{pmatrix}
\begin{pmatrix}
c_{A,k}
\\
c_{B,k}
\end{pmatrix}
=E(k)
\begin{pmatrix}
1&g(k)
\\
g^*(k)&1
\end{pmatrix}
\begin{pmatrix}
c_{A,k}
\\
c_{B,k}
\end{pmatrix},
\end{equation}
where the nearest-neighbour off-diagonal matrix elements are $h(k)=t+t'e^{-ikd}$ and $g(k)=\eta+\eta'e^{-ikd}$. The detail of the on-diagonal matrix element $\epsilon$ is unimportant, however it may not be trivially ignored at this point due to the presence of non-zero off-diagonal elements within $S(k)$.

In fact, although it is possible to eliminate it from the diagonal of Eq.~(\ref{eqn:TBSE}) by redefining the zero of the energy, $E(k) = \epsilon + {\tilde E}(k)$, it reappers in the off-diagonal elements. Upon such redefinition, $h(k)$ becomes $h(k) - \epsilon g(k)$, while $S(k)$ retains its off-diagonal elements, and so its elimination in this manner is fruitless.

The analytic calculation of all the tight-binding parameters $\epsilon,t,t',\eta,\eta'$ may be found in App.~\ref{app:B} alongside a discussion with respect to the inherent `gauge ambiguity' present when defining the phases of the hopping parameters. The validity of the tight-binding approximation guarantees that $\epsilon,t,t'<0$ and $0\leq\eta,\eta'\leq1$.

The convention in most tight-binding approximations is to simply ignore the overlap matrix, setting $\eta,\eta'=0$. This is because it either does not contribute meaningfully due to its negligibility~\cite{Asboth:2016,SSH:1979,SSH:1980}, or it has no effect on the interesting low-energy physics\cite{Reich:2002,Sandu:2005}. However, in most candidate systems where non-Hermitian topology could be observed, for example plasmonic lattices\cite{Poddubny:2014,Downing:2017,Downing:2018,Kruk:2017,Pocock:2018,Yousefi:2019,Wang:2016}, neither of these situations is in principle realised due to the long-ranged natures of the interactions.

In the next section, we show that when the overlap matrix is not approximated with the identity matrix, {\it i.e.} $g(k)\neq0$, then the edge states display edge polarisation and appear attenuated and amplified. To do so, we solve the eigenvalue problem for a finite system obtained by means of a simple second quantisation procedure applied to the tight-binding problem of Eq.~(\ref{eqn:TBSE}). We show results for two distinct cases, in which either the distances between Dirac delta functions, $v$ and $w$, or baseline potentials, $V_v$ and $V_w$, are varied. In both cases, edge polarisation is observed.

In the subsequent section, by investigating the bulk system, such effects are shown to be akin to those found in non-Hermitian systems. In fact, when forced to describe an eigenvalue problem in the same sublattice basis used to solve the finite chain, the effective Hamiltonian assumes a non-Hermitian form. This is due to the presence of the overlap matrix, which effectively introduces next-nearest-neighbour interactions that manifest as balanced gains and losses. The topology of the Brillouin zone is shown, however, to be identical to that of the standard Hermitian SSH model. Therefore, the edge states are still protected by chiral symmetry.

\section{The Finite Solution}\label{sec:finite}

To solve the finite system we go beyond first quantisation, wherein the tight-binding parameters were determined as in App.~\ref{app:B}, and extend to a second quantisation in terms of creation and annihilation operators. To do so, we consider the Schr{\"o}dinger equation within the bulk as initially defined in Eq.~(\ref{eqn:TBSE}) and postulate the second quantised full-chain Hamiltonian and overlap operators that would generate this bulk equation upon the imposition of periodic boundary conditions, {\it i.e.} Bloch's theorem.

Such a postulation is a simple task since only nearest-neighbour interactions are considered. So $\hat{H}\ket{0}=E\hat{S}\ket{0}$ is found where $\ket{0}$ is the vacuum state and the Hamiltonian and overlap operators are expressed in terms of lattice-site creation and annihilation operators, $\hat{c}_{\alpha, i}^\dagger$ and $\hat{c}^{}_{\alpha, i}$ respectively, as:
\begin{equation}\label{eqn:Hop}
\hat{H}=\epsilon{\hat H}_{\rm os} +t{\hat H}_{\rm R} +t'{\hat H}_{\rm L},~~
\hat{S} = {\hat H}_{\rm os} +\eta{\hat H}_{\rm R} +\eta'{\hat H}_{\rm L},
\end{equation}
where ${\hat H}_{\rm os} = \sum_i (\hat{c}^\dagger_{A,i}\hat{c}^{}_{A,i} +\hat{c}^\dagger_{B,i}\hat{c}^{}_{B,i})$, ${\hat H}_{\rm R} =  \sum_i (\hat{c}^\dagger_{A,i}\hat{c}^{}_{B,i}+\hat{c}^\dagger_{B,i}\hat{c}^{}_{A,i})$, and ${\hat H}_{\rm L} =  \sum_i (\hat{c}^\dagger_{A(i+1)}\hat{c}^{}_{Bi}+\hat{c}^\dagger_{Bi}\hat{c}^{}_{A(i+1)})$. Here $i$ denotes the unit cell and $\alpha=A,B$ the sites within it.

Then, the matrix eigenvalue equation $\hat{H}\ket{0}=E\hat{S}\ket{0}$ must be solved numerically for an arbitrary number of unit cells. We expect edge and defect states to occur at the junction of two separate chains that have differing bulk topological invariants. At the edges the chain terminates with the vacuum and, since the vacuum always has a trivial bulk invariant, if the unit cell is topologically non-trivial then edge states will exist.

Finally, due to the overlap between neighbouring sites, the on-site potentials of the lattice sites may not be ignored. Within the bulk, these are simply $\epsilon$ since there are Dirac-delta potentials that neighbour each site on either side. However, at the edges, there are only potentials in one direction (that which is opposite to the vacuum) and so the on-site potentials at the edges, denoted $\varepsilon$, are different to, but crucially smaller in magnitude than, those within the bulk.

As a result, the edge states will never be forbidden by this potential nor is the observed edge polarisation caused by this fact. This final point is due to the fact that changing $\varepsilon$ by altering the boundary conditions does not lift the degeneracy of the edge modes nor modify the character of their wavefunctions. Its only effect is to make the edge state energy non-flat in a similar way as seen in Ref.~\onlinecite{Smith:2019} where, in hindsight, a poor choice of boundary condition was also made.

We now consider two separate scenarios.
In the first, the distances between the Dirac-delta potentials are modulated with the baseline potentials kept constant and equal to zero\cite{Smith:2019}. In the second, the distances between potentials are held constant, whilst the baseline potentials between them are varied. Both reduce, in the tight-binding limit, to the usual SSH model. However, in the latter case, there is a need to consider longer-range interactions when the baseline potentials begin to differ significantly. On the other hand, in the former case, this is not a requirement and so it is a better demonstration of the observed phenomenon.

To keep the presentation compact, we will only discuss the former in the main text with the latter analysed in App.~\ref{app:baseline}. Although the two differ in the details described above, the main features discussed here, {\it i.e.} the  attenuation and amplification of topologically protected edge states, are common to both. This is a testament to the topological nature of such behaviour, that does not depend on details in the construction of the model. Furthermore, in an effort to maintain clarity within the prose, all hopping parameters are quoted within App.~\ref{app:B} since they are unimportant to the narrative of the work.

In this former case, the tight-binding parameters of Eq.~(\ref{eqn:hopparams}) simplify considerably to those given in Eq.~(\ref{eqn:prosehopparams}) where $V$ is the potential of each Dirac-delta such that $E_0=-mV^2/(2\hbar^2)$ is the energy of a lone Dirac-delta potential. We take natural units of $\hbar=m=1$, a Dirac-delta strength of $V=-10$, and vary the distances between the Dirac-deltas as $v=a$ and $w=d-a$ such that the unit-cell length, which will be taken to be $d=1$, remains constant.

Considering the boundary conditions at the edges, there is only one nearest-neighbour for the boundary site and so the tight-binding on-site potentials therein are
$\varepsilon_v=E_0[1+2(e^{-2\kappa v}+2e^{-2\kappa d})]$ and $\varepsilon_w=E_0[1+2(e^{-2\kappa w}+2e^{-2\kappa d})]$, where $\varepsilon_v$ and $\varepsilon_w$ apply if the final hopping is of the $v$-type, $t$, or $w$-type, $t'$, respectively. Comparing these to that of the bulk, $\epsilon=E_0[1+2(e^{-2\kappa v}+e^{-2\kappa w}+2e^{-2\kappa d})]$, from Eq.~(\ref{eqn:prosehopparams}), shows that the boundary potentials are always lesser in magnitude than the the bulk potentials.

\begin{figure}
\centering
\begin{minipage}{.5\linewidth}
\begin{overpic}[width=\linewidth]{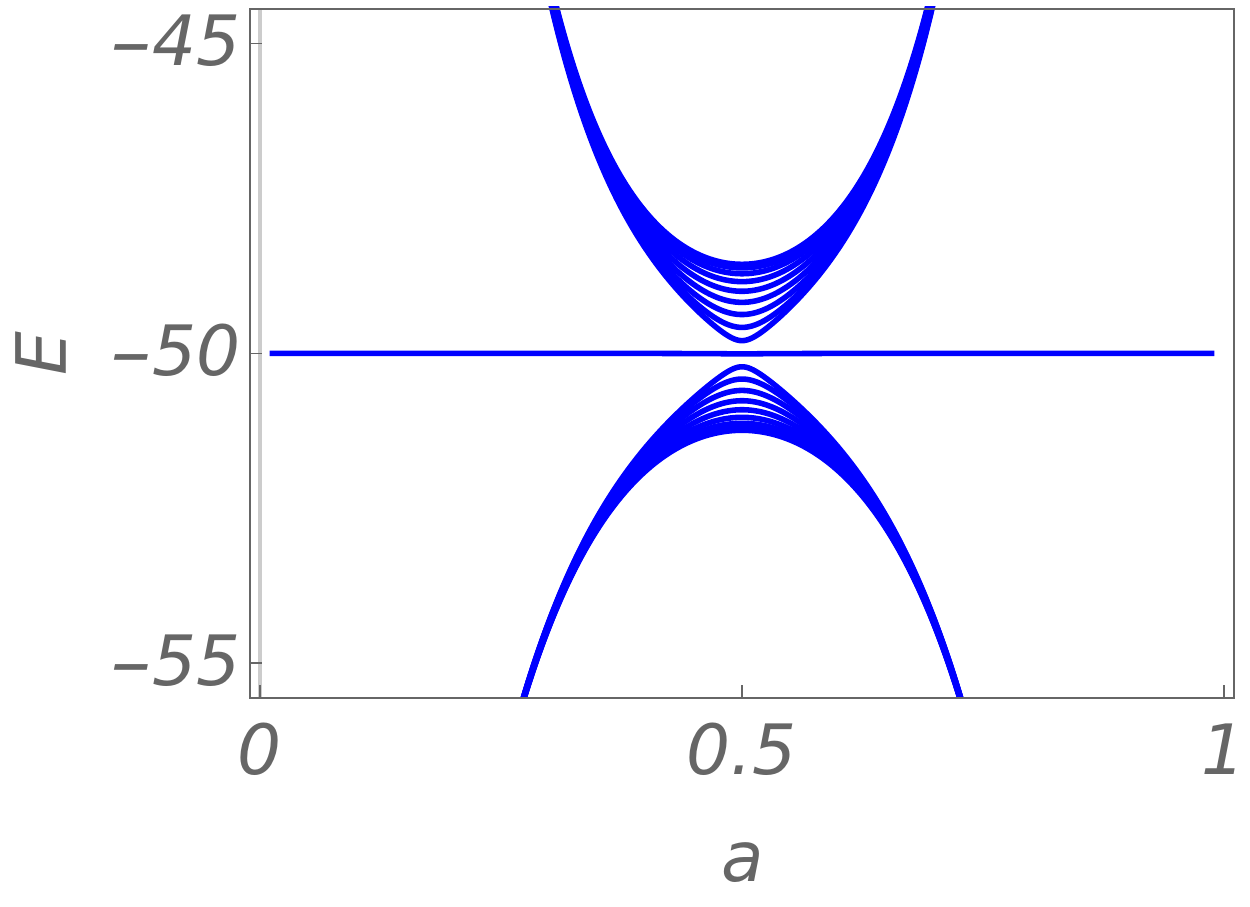}\put(22,65){(a)}
\end{overpic}
\begin{overpic}[width=\linewidth]{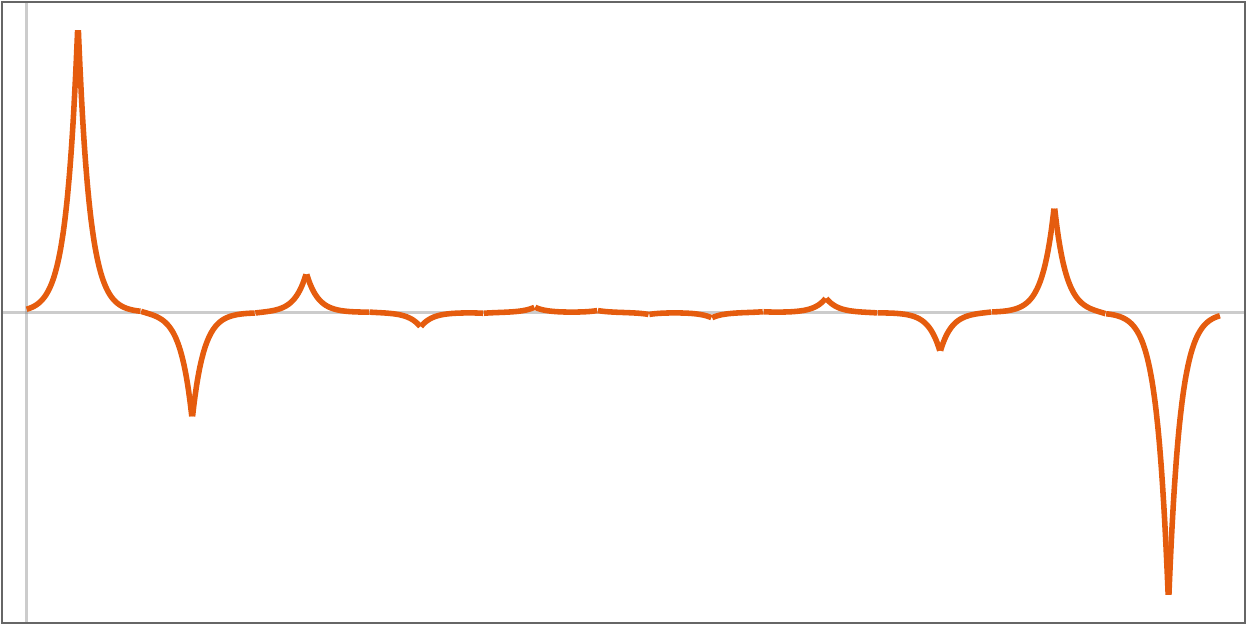}\put(88,43){(c)}
\end{overpic}
\begin{overpic}[width=\linewidth]{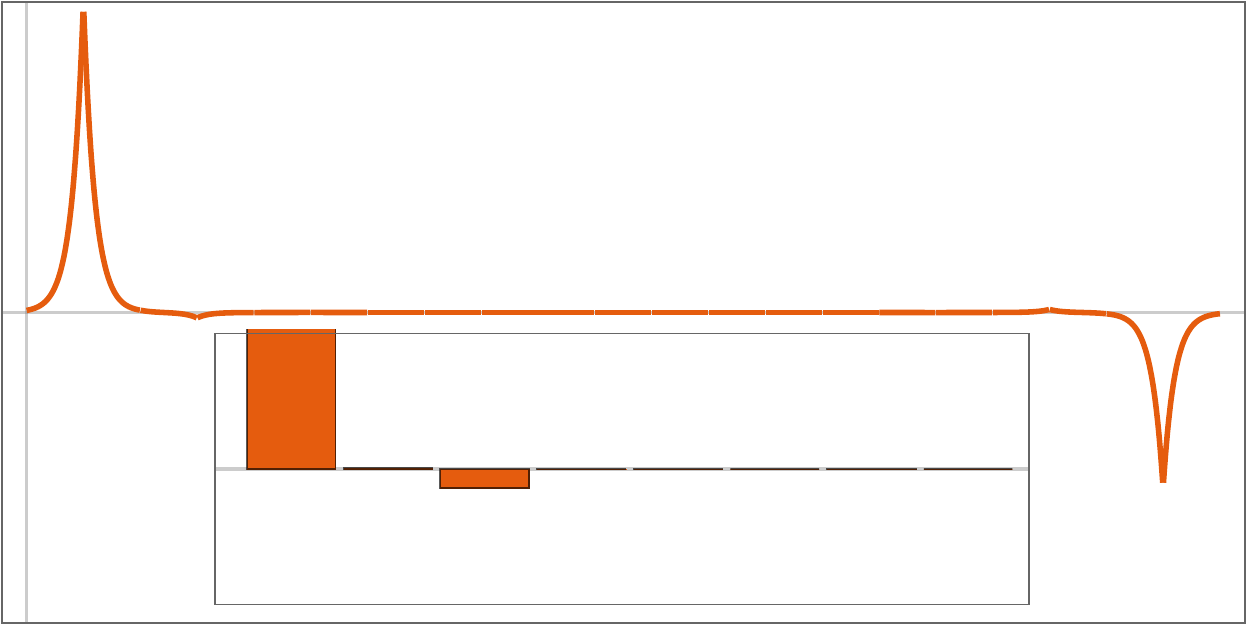}\put(88,4){(e)}
\end{overpic}
\begin{overpic}[width=\linewidth]{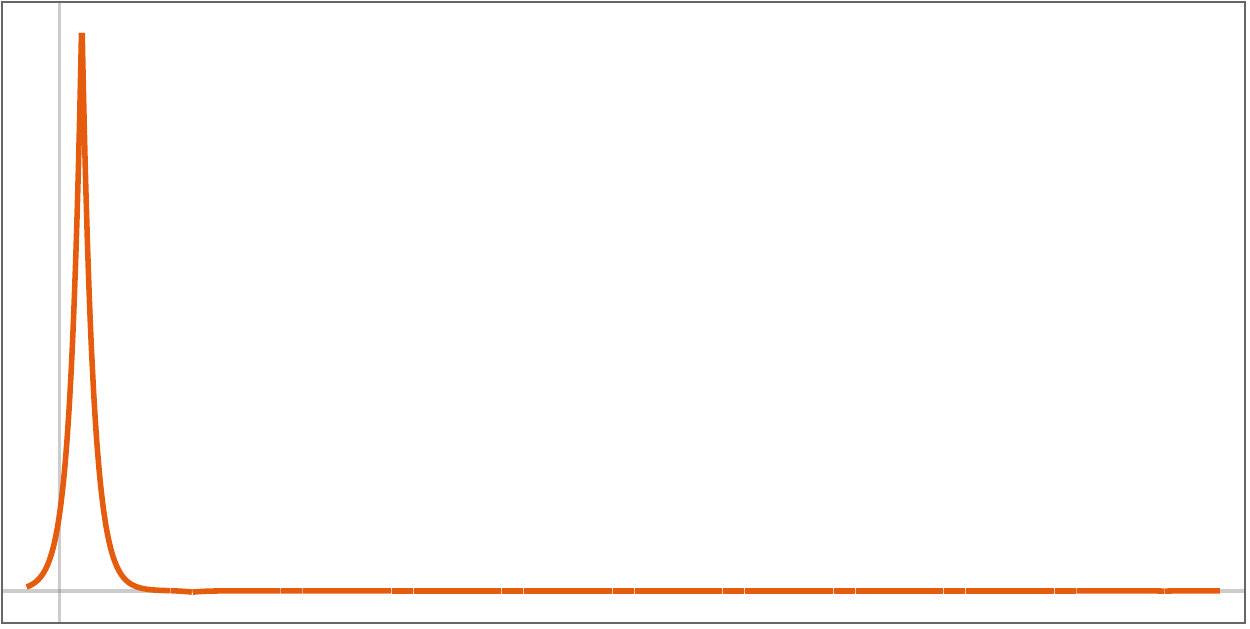}\put(88,41){(g)}
\end{overpic}
\end{minipage}%
\begin{minipage}{.5\linewidth}
\begin{overpic}[width=\linewidth]{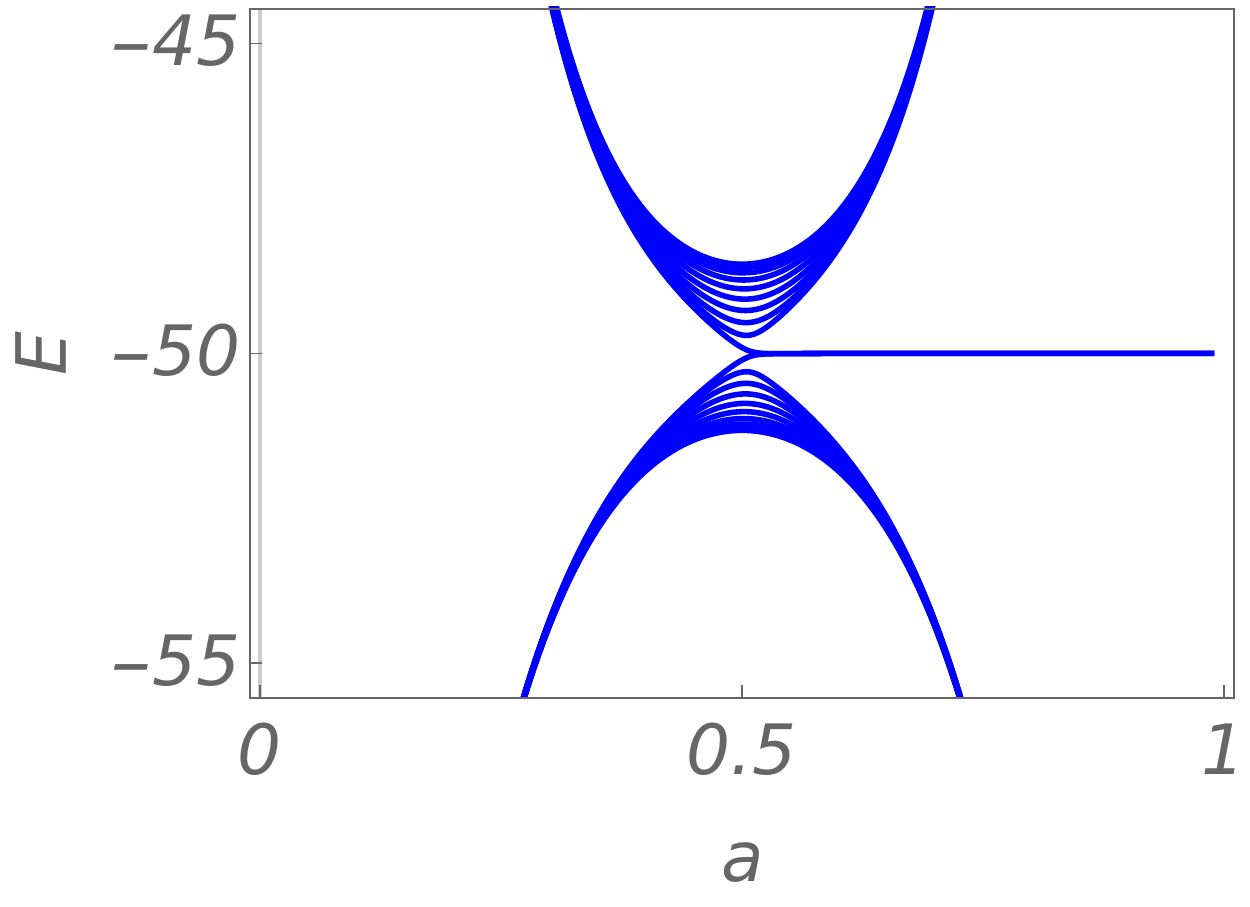}\put(22,65){(b)}
\end{overpic}
\begin{overpic}[width=\linewidth]{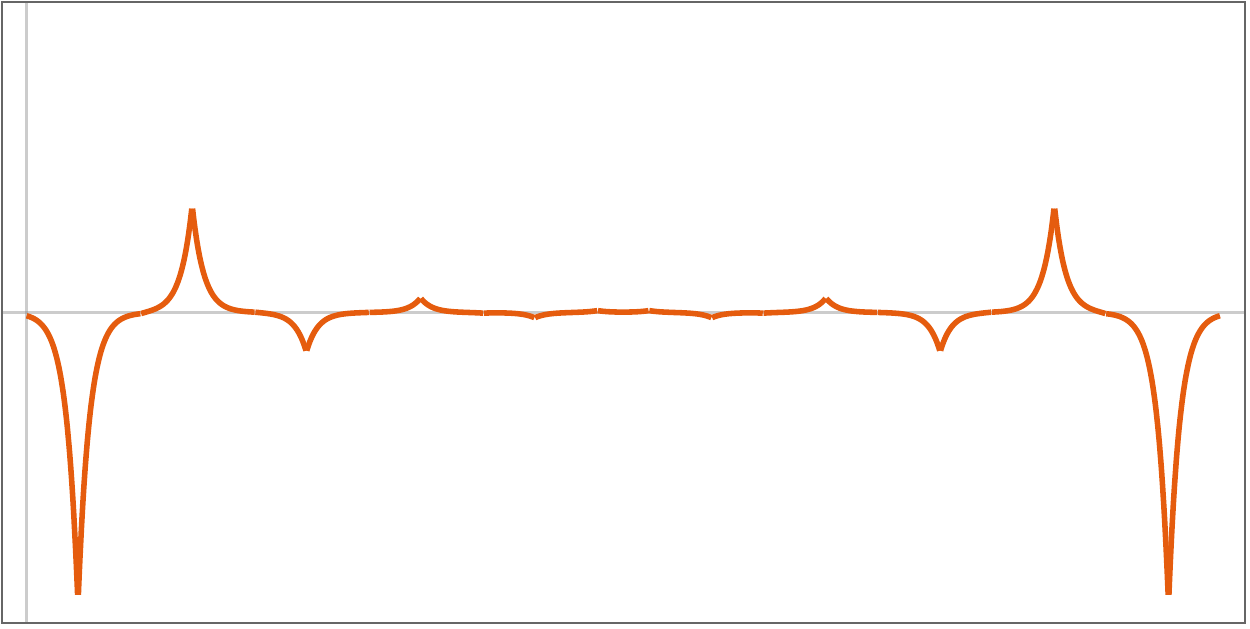}\put(6,43){(d)}
\end{overpic}
\begin{overpic}[width=\linewidth]{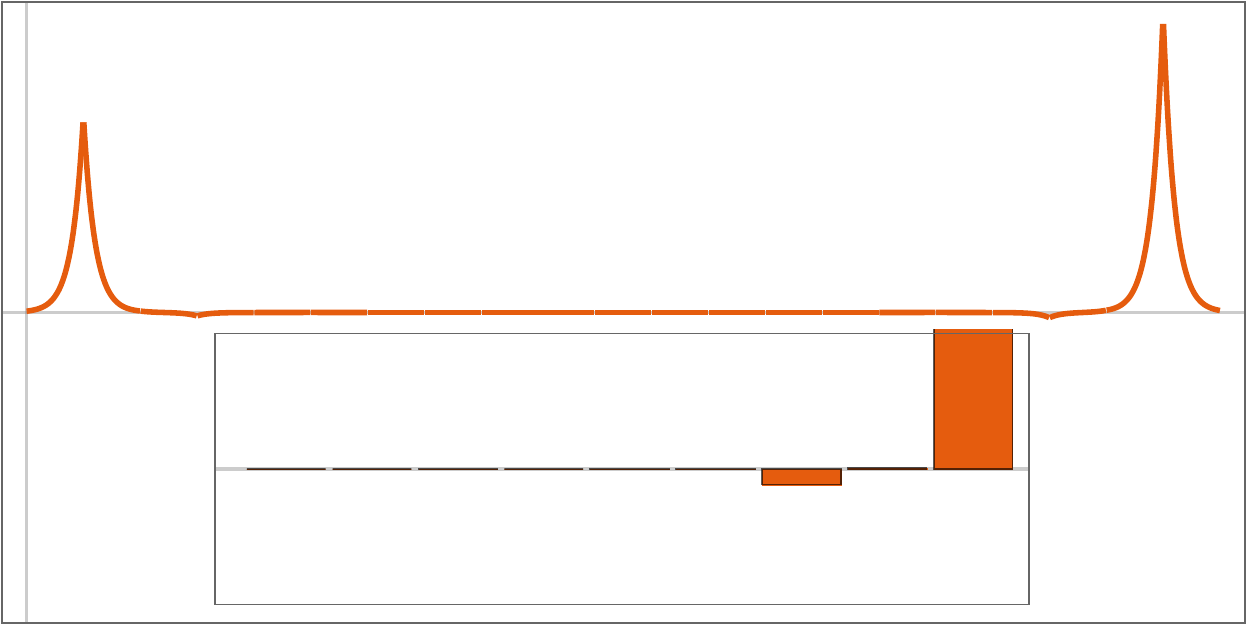}\put(6,4){(f)}
\end{overpic}
\begin{overpic}[width=\linewidth]{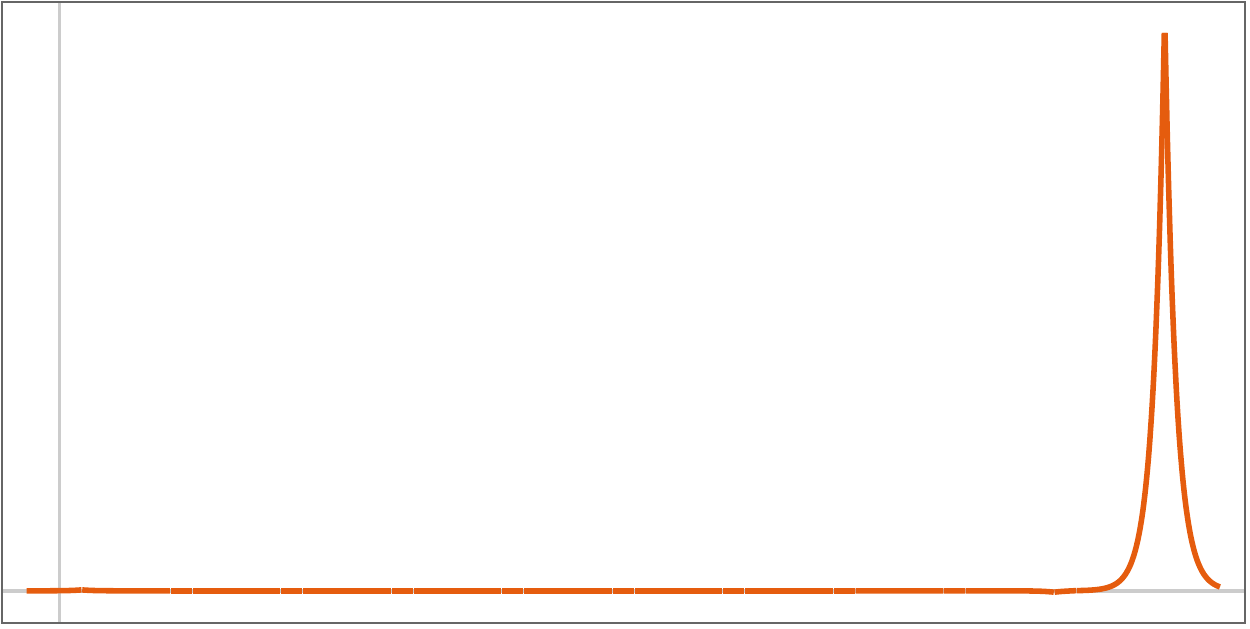}\put(6,41){(h)}
\end{overpic}
\end{minipage}
\caption{(Colour on-line) A selection of figures applying to the system wherein only the distances between the Dirac-deltas are varied. Panels (a,b): the finite band spectrum for $N=19$ (a) and $N=20$ (b) lattice sites with $v=a$, $w=1-a$. Panels (c,d): the edge state wavefunctions for $a=0.55$ corresponding to the mid-gap modes of panel (b). Panels (e,f): the same edge state wavefunctions for $a=0.7$. The insets reveal the lattice-site weights $c_A,c_B$ close to the edge. Panels (g,h): again, the same edge state wavefunctions for $a=0.8$.}
\label{fig:first}
\end{figure}

This boundary condition generates the spectra shown in Fig.~\ref{fig:first}(a,b) and corresponds to `open' boundary conditions, whereby the states are allowed to decay to spatial infinity at the edges. As may be seen, there exist mid-gap edge states in both cases of $N=19$ (with the final sites on the same sublattice) and $N=20$ (with the final sites on opposite sublattices) that are seemingly identical to the standard SSH solution~\cite{Asboth:2016}.

In Figs.~\ref{fig:first}(c,d) we show the wavefunctions of edge states for a chain with $N=20$ sites when $a=0.55$. As in the conventional SSH model, they appear confined to a single sublattice only, {\it i.e.} the wavefunction vanishes on all $B$ sites. This fact confirms that such states are eigenstates of the chiral operator $\sigma_z$ and are thus protected by the presence of this symmetry~\cite{Asboth:2016}. Note that the wavefunctions have the same weight on the outer sites. As the asymmetry parameter $a$ is increased, the edge state wavefunctions change dramatically.

As shown in Figs.~\ref{fig:first}(e,f), when $a=0.7$ they appear to be amplified at one edge and attenuated at the other. In the extreme case of $a= 0.8$, in Figs.~\ref{fig:first}(g,h), they appear completely localised at one edge. To illustrate the point, it must be reiterated that if this system were solved (incorrectly) with $\varepsilon_{v,w}=\epsilon$ then these states would have the same character as in panels (c-h). Therefore the boundary conditions do not cause the edge polarisation.

In addition, by observing the insets of panels (e,f), it may be noted that, for intermediate values of $a$, the edge states are {\it always} localised on one sublattice only. Hence chiral symmetry is never broken. This striking result shows that the edge states are still topologically protected, since to have weights on only a single sublattice requires that they be eigenstates of the chiral symmetry operator\cite{Asboth:2016}, and therefore their localisation at either edge cannot stem from a (trivial) breaking of the chiral symmetry protecting them. This effect is rather due to the (usually neglected) overlap matrix. In the next section, we show that such an overlap can be used to give rise to an {\it effective} non-Hermitian ${\cal PT}$-symmetric~\cite{Kawabata1:2019,Yuce:2019} tight-binding eigenvalue problem, which explains the observed edge polarisation typical of such non-Hermitian problems.

\section{The Bulk Solution}\label{sec:bulk}

To obtain an eigenvalue problem in the conventional form of ${\cal H}(k){\bar \psi}_k=E(k){\bar \psi}_k$, it is standard practice to multiply Eq.~\eqref{eqn:TBSE} through by $S^{-1/2}(k)$. When done so, one identifies ${\cal H}(k) \equiv S^{-1/2}(k)H(k)S^{-1/2}(k)$ as an Hermitian matrix, and $\bar{\psi}_k \equiv S^{1/2}(k)\psi_k$ as the corresponding transformed wavefunction.

Such a wavefunction is expressed in a basis which mixes the sublattices and (in an open-chain setting) combines each site with its nearest neighbours, second-nearest neighbours, and so on. Therefore, in such a case, it would be not only difficult but also needlessly obscuring to compare the open chain (which is expressed in the original basis of lattice sites) with the periodic one. This would also obscure the requisite bulk-boundary correspondence.

For this reason, we now stray from the conventional wisdom and define the eigenvalue problem by multiplying Eq.~(\ref{eqn:TBSE}) through by $S^{-1}(k)$ rather than $S^{-1/2}(k)$. The resulting Hamiltonian ${\cal H}(k) \equiv S^{-1}(k)H(k)$ is now {\it non-Hermitian}:
\begin{equation}\label{eqn:NHHam}
\begin{aligned}
{\cal H}(k)
&=
\begin{pmatrix}
\varepsilon(k)+i\gamma(k)&f(k)
\\
f^*(k)&\varepsilon(k)-i\gamma(k)
\end{pmatrix},
\end{aligned}
\end{equation}
where:
\begin{equation}\label{eqn:bulkparams}
\begin{aligned}
\varepsilon(k)&=[\epsilon-t\eta-t'\eta'-(t\eta'+t'\eta)\cos(kd)]/(1-|g(k)|^2),
\\
f(k)&=[h(k)-\epsilon g(k)]/(1-|g(k)|^2),
\\
\gamma(k)&=[t\eta'-t'\eta]\sin(kd)/(1-|g(k)|^2).
\end{aligned}
\end{equation}
We note that the Hamiltonian in Eq.~(\ref{eqn:NHHam}) has a form identical to the one studied in Ref.~\onlinecite{Yuce2:2018}.
The multiplication by the overlap matrix introduces {\it imaginary} diagonal terms, $\pm i\gamma(k)$, which resemble compensated gains and losses in next-nearest-neighbour hopping processes. The fact that such terms have a dependence on the wavevector $k$ is required by the fact that the matrix multiplication by $S^{-1}(k)$ should not change the topological class to which the Hamiltonian belongs to.

We stress that the non-Hermiticity of Eq.~(\ref{eqn:TBSE}) is an artificial mathematical feature. It is the price we pay to obtain a conventional eigenvalue problem that is still expressed in the original sublattice basis of $\psi_k=(c_A,c_B)^{\rm T}$. This is essential in order to define a one-to-one mapping between the Hamiltonians of the periodic and finite systems. Only in this case can we in fact meaningfully discuss the topological aspects of the bulk in relation to the finite system through the bulk-boundary correspondence. For this, it is required that the Hilbert spaces of the two systems can be mapped into each other~\cite{Asboth:2016,Huang:2018}.

However, the non-Hermiticity also enables us to readily give meaning to the observed physical behaviours of the edge modes that remain topologically protected and degenerate yet display attenuation and amplification. It is crucial to note that, as shown in the previous section, such edge effects arise {\it independently} of the transformation adopted for the periodic system but acquire a simple interpretation when the matrix is made non-Hermitian by forcing the eigenvalue problem to be expressed in the same basis.

The low-energy effective Hamiltonian may now be decomposed in terms of the Pauli matrices $\bm{\sigma}=(\sigma_x,\sigma_y,\sigma_z)$ as:
${\cal H}(k)=d_0(k)\mathbb{1}_2+\bm{d}(k)\cdot\bm{\sigma}$,
where $d_0(k)=\varepsilon(k)$ and $\bm{d}(k)=(d_x(k),d_y(k),d_z(k))$ with $d_x(k)={\rm Re}[f(k)]$, $d_y(k)={\rm Im}[f(k)]$ and $d_z(k)=i\gamma(k)$.
In the conventional SSH Hermitian system, $d_z=0$ and so the topological invariant that protects the edge states via the presence of chiral symmetry resides within the winding of $\bm{d}=(d_x,d_y,0)$. This also applies within non-Hermitian systems wherein $d_z\neq0$ as long as it is possible to adiabatically deform the two systems to one another. As such, the $d_0(k)\mathbb{1}_2$ is not important with respect to the topological information and symmetry protection. Indeed, its contribution to the eigenvalues may be trivially removed without affecting the underlying topology of the Hamiltonian\cite{Li:2019}.

Solving the eigenvalue problem presented through Eq.~(\ref{eqn:NHHam}), we find that:
\begin{equation}
E_\pm(k)=\frac{\varepsilon(k)\pm\sqrt{|f(k)|^2-\gamma^2(k)}}{1-|g(k)|^2},
\end{equation}
and so we have complex energy solutions if $\gamma(k)>|f(k)|$ at any point within the Brillouin zone where the exceptional points are defined by $|f(k)|=\gamma(k)$. This is never achieved within physical tight-binding models, however, as it ought not to. A proof of this for the present model is shown in App~\ref{app:C}. Since $\gamma(k)\propto\sin(kd)$, the exceptional points annihilate one another as they coincide with the topological transition point\cite{Yuce2:2018} whereat the band gap at the edge of the Brillouin zone, $k=\pm\pi/d$, closes.

These results follow naturally since the original Hamiltonian of our problem is demonstrably Hermitian in nature. It is only the {\it effective} tight-binding Hamiltonian that has a non-Hermitian character. Moreover, this non-Hermitian character enters in since we may not neglect the overlap of the neighbouring basis wavefunctions. As such, we say that this effective Hamiltonian is ${\cal PT}$-symmetric since it possesses real eigenvalues in the absence of Hermiticity\cite{Yuce:2019}. Yet this {\it effective} non-Hermiticity has a highly non-trivial impact on the behaviour of the edge states. The latter is a physical effect of the system and not an artifact of the tight-binding approximation, the exact solution to the problem exhibits this same phenomenon\cite{Smith:2019}, nor of the used transformation involving $S^{-1}(k)$ since this transformation is not manifest in the finite system.

Since $|f(k)|^2-\gamma^2(k)>0$ always, as above, the present class of Hamiltonian may be adiabatically deformed into its Hermitian counterpart. This is because, as long as the band gap does not close in such a deformation, then the exceptional points are never encountered (recalling that they reside at the transition point). Then, through this adiabatic deformation, the $i\gamma(k)\sigma_z$ term may be eliminated from the tight-binding Hamiltonian. Hence the topological character is that of the basic SSH model, {\it i.e.} non-trivial, and is given by the $\mathbb{Z}$-invariant that resides in the winding number of the off-diagonal element $f(k)$ (or equivalently the Zak phase $\theta_{\cal Z}$).

This does not mean to say that $\gamma(k)$ is irrelevant. Only that it does not affect the topological nature of the Hamiltonian, and therefore the protection of the edge states. Its effect is in fact to alter the physical character of such topological modes through the phenomenon of edge polarisation and not their underlying topological nature.

In fact, we can identify this Hamiltonian as belonging to the BDI class that is characterised by the following symmetries: particle-hole (PHS), time-reversal (TRS), and chiral (CS). Mathematically, these are expressed as\cite{Kawabata1:2019}:
\begin{equation}
\begin{aligned}
{\rm PHS}:~&\hat{{\cal C}}_-^{-1}{\cal H}^{\rm T}(k)\hat{{\cal C}}_-=-{\cal H}(-k),
\\
{\rm TRS}:~&\hat{{\cal T}}_+^{-1}{\cal H}^*(k)\hat{{\cal T}}_+={\cal H}(-k),
\\
{\rm CS}:~&\hat{\Gamma}^{-1}{\cal H}^\dagger(k)\hat{\Gamma}=-{\cal H}(k),
\end{aligned}
\end{equation}
where the operators may be found in the present case to be: $\hat{{\cal C}}_-=\sigma_z$, $\hat{{\cal T}}_+=\mathbb{1}_2$, and $\hat{\Gamma}=\hat{{\cal C}}_-\hat{{\cal T}}_+=\sigma_z$. As such, the Hamiltonian possesses chiral symmetry guaranteed by $\hat{\Gamma}=\sigma_z$. However, and this is the crucial difference with Hermitian problems, the chiral and sublattice symmetries are not identical. The definition of sublattice symmetry is\cite{Kawabata1:2019}:
\begin{equation}
{\rm SLS}:~\hat{{\cal S}}^{-1}{\cal H}(k)\hat{{\cal S}}=-{\cal H}(k),
\end{equation}
which may be shown to be present for the current effective Hamiltonian with $\hat{{\cal S}}=\sigma_z-i\gamma(k)(\sigma_x-i\sigma_y)/f(k)$.

It is clear then that these two symmetries (chiral and sublattice) coincide in the Hermitian limit wherein $\gamma(k)\to0$. It is interesting to note that this sublattice symmetry operator acts upon the wavefunctions to mix their weights on different sublattices in a $k$-dependent fashion. The latter implies that, in an open-chain setting, the transformation must mix a site with its nearest neighbours, next-nearest neighbours, and so on.

Since the effective Hamiltonian ${\cal H}(k)$ belongs to the BDI class, according to the periodic table of invariants\cite{Gong:2018} we seek a $\mathbb{Z}\oplus\mathbb{Z}$ invariant, {\it i.e.} the sum of two separate and distinct invariants.
These invariants are the parametric winding numbers, in the complex plane, of the off-diagonal matrix element\cite{Asboth:2016}, $f(k)$, and the imaginary part of the energy eigenvalue\cite{Yuce:2019,Song:2019}. Since the energy is always real, this second winding number is identically zero and so there is only one $\mathbb{Z}$-invariant: the winding number of the off-diagonal element of the Hamiltonian. This has the following simple analytical form of:
\begin{equation}
{\cal W}=\frac{1}{2\pi i}\int_{-\pi/d}^{+\pi/d}dk\frac{d}{dk}\ln[f(k)]=
\begin{cases}
1,\quad a>d/2,
\\
0,\quad a<d/2.
\end{cases}
\end{equation}
In the presence of chiral symmetry, this invariant is identical to the (non-Hermitian) Zak phase\cite{Ghatak:2019,Asboth:2016} given in Eq.~(\ref{eq:non_herm_zak}), which reduces to that given in Eq.~(\ref{eq:herm_zak}) in the presence of ${\cal PT}$-symmetry\cite{Zhang:2019}.

A crucial point to make here, which highlights the broader scope of this paper, is that the analysis of the topological character of the above Hamiltonian is in fact completely general. It does not only apply to the present Kronig-Penney system. Given a tight-binding Schr{\"o}dinger equation as in Eq.~(\ref{eqn:TBSE}), the effective Hamiltonian can always be made non-Hermitian and ${\cal PT}$-symmetric as in Eq.~(\ref{eqn:NHHam}). This then allows to connect the described bulk effects to the observed edge effects through bulk-boundary correspondence. Therefore, provided that it is not completely negligible, the overlap matrix within all tight-binding models will cause edge polarisation by attenuating and amplifying any edge states observed in the system.

\section{Summary and Conclusion}
We have shown that the overlap between neighbouring lattice sites, often ignored for simplicity in tight-binding problems, has a non-trivial influence on the behaviour and nature of the topological edge states in the finite system. The observed attenuation and amplification of such states can be interpreted in terms of an {\it effective} non-Hermitian edge polarisation. To make this apparent within the periodic system, we have performed a non-standard transformation of the generalised eigenvalue problem that keeps it in the same sublattice basis. The resulting conventional eigenvalue problem is non-Hermitian, which is the price we pay for fixating with the sublattice basis.

Instead of being a drawback, the non-Hermiticity of the effective problem allows for our recognition of the topological origin of the observed edge polarisation. 
The effective non-Hermitian Hamiltonian belongs to the BDI class of topological invariants~\cite{Gong:2018,Kawabata1:2019}. As such, the sought invariant is $\mathbb{Z}\oplus\mathbb{Z}$, composed by the winding number, ${\cal W}$, of the off-diagonal matrix element of the Hamiltonian (or equivalently the Zak phase $\theta_{\cal Z}$), and the winding number of the imaginary part of the energy eigenvalue. Since the original system is Hermitian, the energy eigenvalues are guaranteed to be real. As such, the derived effective Hamiltonian is ${\cal PT}$-symmetric; this allows it to be adiabatically deformed to its Hermitian counterpart, from which it inherits the topology. Therefore, the $\mathbb{Z}$-invariant is given by the winding number of the off-diagonal matrix element, ${\cal W}$, or (equivalently in the presence of chiral symmetry) the Zak phase $\theta_{\cal Z}$.

Our result shows that edge polarisation alone is not sufficient as an experimental diagnostic of non-Hermitian topology since it can equally appear in Hermitian problems as the one discussed here. This is particularly crucial for, e.g., plasmons in metal gratings or nanoparticle arrays, one of the possible test-beds of non-Hermitian topology, and to which the present model applies~\cite{Smith:2020,DellaValle:2010,Chaves:2019}.

Although in many electronic systems the `atomic' (lattice-site) wavefunctions are strongly localised within low-lying orbitals allowing the overlap matrix to be safely ignored, this is not the case in all contexts. In fact, the topological protection of edge states is not unique to electronic systems. Indeed, since it is a feature of wave-like excitations (since the topological character may be found within the wavefunction through the Zak or Berry phase), symmetry protected states may be observed in photonic\cite{Liu:2018,Wang1:2018,Ozawa:2019,Gorlach:2017}, phononic\cite{Pal:2018,Zhao:2018,Zhang:2018}, magnonic\cite{Mei:2019,Qin:2017,Pirmoradian:2018}, and plasmonic\cite{Poddubny:2014,Downing:2017,Downing:2018,Kruk:2017,Pocock:2018,Yousefi:2019,Wang:2016} systems.

In such systems, especially the photonic and plasmonic ones, the interactions are often long-ranged and any localised states, with which a tight-binding model may be constructed, have evanescent tails far from their lattice sites. Then the overlap matrix would be important to account for. Especially if the tight-binding model is not found through first-principles but instead through some {\it ad hoc} fitting procedure. In such a case, an accidental ignorance of the overlap matrix could limit the predictive power of the constructed model.

\section{Acknowledgements}

T.B.S. acknowledges the support of the EPSRC Ph.D. studentship grant EP/N509565/1. A.P. and T.B.S. acknowledge support from the Royal Society International Exchange grant IES\textbackslash R3\textbackslash 170252.

\bibliography{bibliography}
\bibliographystyle{apsrev4-1}

\appendix
\onecolumngrid

\section{The Solution to the Solitary Asymmetric Dirac-delta\label{app:A}}

For a bipartite Kronig-Penney model with Dirac-delta potentials that have negative strengths we seek negative energy solutions. A lone Dirac-delta potential possesses a single bound state that exponentially localises to the potential site.

The solution of the lone asymmetric bound state proceeds in the standard scattering way. Assuming the Dirac-delta potential to be situated at $x=x_0$, we solve:
\begin{equation}\label{eqn:TISE}
H(x)\Psi(x)=\left[-\frac{\hbar^2}{2m}\frac{d^2}{dx^2}+V(x)\right]\Psi(x)=E\Psi(x),
\end{equation}
where $V(x)=V\delta(x-x_0)+V_1\theta(x_0-x)+V_2\theta(x-x_0)$, with the standard solution for the wavefunction of:
\begin{equation}
\Psi(x,x_0)=\theta(x_0-x)(Ae^{iq_1x}+Be^{-iq_1x})
+\theta(x-x_0)(Ce^{iq_2x}+De^{-iq_2x}),
\end{equation}
and seek negative energy solutions such that $q_j=i\kappa_j$, where these wavevectors are found by solving the TISE in each region separately as:
\begin{equation}\label{eqn:wavevectors}
q_j=\hbar^{-1}\sqrt{2m(E-V_j)}\implies \kappa_j=\hbar^{-1}\sqrt{2m(V_j-E)}.
\end{equation}
As a result, the wavefunction must be well-defined at $\pm\infty$ and thus $A=D=0$ thereby yielding:
\begin{equation}
\Psi(x,x_0)=\theta(x_0-x)Be^{\kappa_1 x}+\theta(x-x_0)Ce^{-\kappa_2 x}.
\end{equation}
Now we enforce the continuity of the wavefunction at $x=x_0$ and so observe that $B=Ce^{-(\kappa_1+\kappa_2)x_0}$. Furthermore, the wavefunction must be normalised as:
\begin{equation}
1=\int_{-\infty}^{+\infty}dx|\Psi(x,x_0)|^2
=C^2e^{-2(\kappa_1+\kappa_2)x_0}\int_{-\infty}^{x_0}dxe^{2\kappa_1 x}+C^2\int_{x_0}^{+\infty}dxe^{-2\kappa_2 x},
\end{equation}
from which it may be seen that: $C=\sqrt{2\kappa_1\kappa_2(\kappa_1+\kappa_2)^{-1}}e^{\kappa_2x_0}$. Thus:
\begin{equation}\label{eqn:BSwf}
\Psi(x,x_0)=\sqrt{\frac{2\kappa_1\kappa_2}{\kappa_1+\kappa_2}}\bigg[\theta(x_0-x)e^{\kappa_1(x-x_0)}
+\theta(x-x_0)e^{-\kappa_2(x-x_0)}\bigg]
\end{equation}

To find the energy of this bound state we integrate the TISE once in the vicinity of the potential since we cannot impose that the derivative of the wavefunction be continuous at the potential due to the presence of the Dirac-delta. Thus:
\begin{equation}
E\int_{x_0-\epsilon}^{x_0+\epsilon}dx\Psi(x,x_0)=-\frac{\hbar^2}{2m}\int_{x_0-\epsilon}^{x_0+\epsilon}dx\Psi''(x,x_0)
+\int_{x_0-\epsilon}^{x_0+\epsilon}dx[V\delta(x)+V_1\theta(x_0-x)+V_2\theta(x-x_0)]\Psi(x,x_0),
\end{equation}
which becomes, in the limit of $\epsilon\rightarrow0$:
\begin{equation}
-\frac{\hbar^2}{2m}\left[\Psi'(x_0^+,x_0)-\Psi'(x_0^-,x_0)\right]+V\Psi(x_0,x_0)
=0,
\end{equation}
where the superscripts $\pm$ signify to take $\Psi'(x,x_0)$ to the limit of $x_0$ within the regions $x>x_0$ (plus) and $x<x_0$ (minus). Thus we see that:
\begin{equation}\label{eqn:divcont}
\frac{\hbar^2}{2m}(\kappa_1+\kappa_2)+V=0,
\end{equation}
and so, taking (\ref{eqn:divcont}) together with (\ref{eqn:wavevectors}), the energy of the bound state, after some unilluminating algebra, is:
\begin{equation}\label{eqn:BSenergy}
E=\frac{1}{2}(V_1+V_2)-\frac{mV^2}{2\hbar^2}-\frac{\hbar^2(V_1-V_2)^2}{8mV^2}.
\end{equation}
Clearly, when $V_1=V_2=0$ we recover the standard result of a symmetric Dirac-delta $E_0=-mV^2/(2\hbar^2)$. Moreover, when $V_1=V_2=U\neq0$ we see that $E=U+E_0$, {\it i.e.} the potentials act as trivial energy shifts; they only have a non-trivial effect when $V_1\neq V_2$.

Taking (\ref{eqn:BSenergy}) and substituting it into the $\kappa_j$ of (\ref{eqn:wavevectors}), yields the wavevectors as:
\begin{equation}
\kappa_{1,2}=\sqrt{\frac{m}{\hbar^2}(V_{1,2}-V_{2,1})+\frac{m^2V^2}{\hbar^4}+\left(\frac{V_1-V_2}{2V}\right)^2}.
\end{equation}
As such, we may see that:
\begin{equation}
\kappa_{1,2}=\sqrt{\left[\frac{V_{1,2}-V_{2,1}}{2V}+\frac{mV}{\hbar^2}\right]^2}
=\pm\left[-\kappa+\frac{1}{2V}(V_{1,2}-V_{2,1})\right],
\end{equation}
where $\kappa=-mV/\hbar^2$, such that $E_0=\kappa V/2$, is the wavevector of a lone symmetric Dirac delta potential. Since it cannot physically be that $\kappa_{1,2}<0$ (such that the correct behaviour at infinity is maintained), the minus sign must be chosen here. As such:
\begin{equation}\label{eqn:BSkappas}
\kappa_1=\kappa-\frac{1}{2V}(V_1-V_2),\quad\kappa_2=\kappa-\frac{1}{2V}(V_2-V_1).
\end{equation}
As may be noted, $\kappa_{1,2}$ can become negative when $\kappa<|(V_2-V_1)(2V)^{-1}|$. In such a case, it would mean that it is no longer energetically favourable for a mode to localise and bind to the Dirac-delta. Instead, it would leak away from it into one of the regions depending on whether $V_2>V_1$ or $V_2<V_1$.

So, in summary, the lone asymmetric Dirac-delta potential hosts a single bound state with negative energy given by (\ref{eqn:BSenergy}) and wavefunction as in (\ref{eqn:BSwf}) where the wavevectors too may be found in (\ref{eqn:BSkappas}). This wavefunction will constitute the atomic orbitals (basis wavefunctions) of our tight-binding model.

\section{The Matrix Elements of the Expansion\label{app:B}}

The full problem, within the bulk as shown in Fig.~\ref{fig:unitcell}, is to solve the time-independent Schr{\"o}dinger equation as in (\ref{eqn:TISE}): $H(x)\Psi(x)=E\Psi(x)$, in the presence of a spatially varying potential given by:
\begin{equation}
V(x)=\sum_i\Big\{V[\delta(x-x_{Ai})+\delta(x-x_{Bi})]
+V_v\theta(x_{Bi}-x)\theta(x-x_{Ai})+V_w\theta(x_{A(i+1)}-x)\theta(x-x_{Bi})\Big\},
\end{equation}
where the sum over $i$ is over a number of unit-cells that is determined by the accuracy required of the model. For small values of $|V|$ and/or $V_{v,w}$, this sum must be over several unit-cells however in the nearest-neighbour approximation it need only be between three neighbouring unit-cells, {\it i.e.} the central unit-cell and its two neighbours.

Considering the unit-cell as shown in Fig.~\ref{fig:unitcell}, the atomic wavefunctions for the $A$ and $B$ sublattices are:
\begin{equation}\label{eqn:basisWFs}
\begin{aligned}
\Psi_A(x,x_{Aj})&=\sqrt{\frac{2\kappa_v\kappa_w}{\kappa_v+\kappa_w}}\left[\theta(x_{Aj}-x)e^{\kappa_w(x-x_{Aj})}
+\theta(x-x_{Aj})e^{-\kappa_v(x-x_{Aj})}\right],
\\
\Psi_B(x,x_{Bj})&=\sqrt{\frac{2\kappa_v\kappa_w}{\kappa_v+\kappa_w}}
\left[\theta(x_{Bj}-x)e^{\kappa_v(x-x_{Bj})}+\theta(x-x_{Bj})e^{-\kappa_w(x-x_{Bj})}\right],
\end{aligned}
\end{equation}
where $\kappa_{v,w}=\kappa-(V_{v,w}-V_{w,v})(2V)^{-1}$. As such, the normalisation constant simplifies to $\mathcal{N}_c=\sqrt{\kappa_v\kappa_w\kappa^{-1}}$. Using these basis wavefunctions, the tight-binding Schr{\"o}dinger equation may be constructed as:
\begin{equation}
\begin{pmatrix}
H_{AA}&H_{AB}
\\
H_{BA}&H_{BB}
\end{pmatrix}
\begin{pmatrix}
c_A
\\
c_B
\end{pmatrix}=E
\begin{pmatrix}
S_{AA}&S_{AB}
\\
S_{BA}&S_{BB}
\end{pmatrix}
\begin{pmatrix}
c_A
\\
c_B
\end{pmatrix},
\end{equation}
where $H_{mn}=\mel{\Psi_m}{H}{\Psi_n}$, $S_{mn}=\bra{\Psi_m}\ket{\Psi_n}$ and $c_n$ are the coefficients that yield the unit-cell wavefunction as $\Psi(x)=\theta(x_{\rm L}-x)\theta(x-x_{\rm R})[c_A\Psi_A(x,x_{A1})+c_B\Psi_B(x,x_{B1})]$. As will be subsequently shown in this section, the relevant tight-binding parameters may be found as:
\begin{equation}\label{eqn:hopparams}
\begin{aligned}
&\epsilon=-\frac{\hbar^2}{2m}\kappa_v\kappa_w\left[1+2\left(e^{-2\kappa_vv}+e^{-2\kappa_ww}
+e^{-2\kappa_vd}+e^{-2\kappa_wd}\right)\right],
\\
&t=-\frac{\hbar^2}{2m}\kappa_v\kappa_w\left[\frac{1}{\kappa}\left(\frac{1}{2\kappa}(\kappa_v^2+\kappa_w^2)+\kappa_v^2v\right)e^{-\kappa_vv}
+e^{-\kappa_vv}+e^{-\kappa_wv}\right],
\\
&t'=-\frac{\hbar^2}{2m}\kappa_v\kappa_w\left[\frac{1}{\kappa}\left(\frac{1}{2\kappa}(\kappa_v^2+\kappa_w^2)+\kappa_w^2w\right)e^{-\kappa_ww}+e^{-\kappa_vw}
+e^{-\kappa_ww}\right],
\\
&\eta=\frac{1}{\kappa^2}\kappa_v\kappa_w\left[\frac{1}{2}\left(e^{-\kappa_vv}+e^{-\kappa_wv}\right)+\kappa ve^{-\kappa_vv}\right],\quad
\eta'=\frac{1}{\kappa^2}\kappa_v\kappa_w\left[\frac{1}{2}\left(e^{-\kappa_vw}+e^{-\kappa_ww}\right)+\kappa we^{-\kappa_ww}\right],
\end{aligned}
\end{equation}
such that $H_{AA}=H_{BB}=\epsilon$, $H_{AB}^{}=H_{BA}^*=te^{ikv}+t'e^{-ikw}$, $S_{AA}=S_{BB}=1$, and $S_{AB}^{}=S_{BA}^*=\eta e^{ikv}+\eta'e^{-ikw}$. In the system as presented within the prose wherein the separations between the Dirac-delta potentials is modified with a constant baseline potential $V_v=V_w=0$, these reduce to:
\begin{equation}
\begin{aligned}\label{eqn:prosehopparams}
\epsilon&=E_0\left[1+2\left(e^{-2\kappa v}+e^{-2\kappa w}+
2e^{-2\kappa d}\right)\right],
\\
t&=E_0\left(3+\kappa v\right)e^{-\kappa v},
\\
t'&=E_0\left(3+\kappa w\right)e^{-\kappa w},
\\
\eta&=\left(1+\kappa v\right)e^{-\kappa v},
\\
\eta'&=\left(1+\kappa w\right)e^{-\kappa w},
\end{aligned}
\end{equation}
where $\kappa=-mV/\hbar^2$ and $E_0=-\hbar^2\kappa^2/(2m)$.

Following the standard, general theory of the tight-binding model, the matrix elements may be evaluated, within the present context as:
\begin{multline}
\mel{\Phi_n}{\hat{H}}{\Phi_m}=\int_{-\infty}^{+\infty}\frac{\kappa dx}{N}\sum_{\{i,j\}=1}^Ne^{ik(x_{mj}-x_{ni})}\left[\theta(x_{ni}-x)e^{\kappa(x-x_{ni})}+\theta(x-x_{ni})e^{-\kappa(x-x_{ni})}\right]\times
\\
\left[-\frac{\hbar^2}{2m}\frac{d^2}{dx^2}+V\sum_{l=-N}^N\left(\delta(x-x_{nl})+\delta(x-x_{ml})\right)\right]\left[\theta(x_{mj}-x)e^{\kappa(x-x_{mj})}+\theta(x-x_{mj})e^{-\kappa(x-x_{mj})}\right],
\end{multline}
\begin{multline}
\bra{\Phi_n}\ket{\Phi_m}=\int_{-\infty}^{+\infty}\frac{\kappa dx}{N}\sum_{\{i,j\}=1}^Ne^{ik(x_{mj}-x_{ni})}\left[\theta(x_{ni}-x)e^{\kappa(x-x_{ni})}+\theta(x-x_{ni})e^{-\kappa(x-x_{ni})}\right]\times
\\
\left[\theta(x_{mj}-x)e^{\kappa(x-x_{mj})}+\theta(x-x_{mj})e^{-\kappa(x-x_{mj})}\right],
\end{multline}
where the sum over $\{i,j\}$ is over pairs of lattice sites, which we restrict to on-site and nearest-neighbours, {\it i.e.} $i,j=0,1,2$, and the sum over $l$ is over a suitable number of Dirac-delta potentials (lattice sites). The prefactor $N$ drops out in the subsequent analysis as it accounts for double counting in the $\{i,j\}$ summation.

Considering first the case wherein $n=m=A$ and $i=j=1$, {\it i.e.} interactions within the unit-cell only, and name this contribution $\epsilon_{AA}$, we see that:
\begin{multline}
\epsilon_{AA}=\frac{2\kappa_w\kappa_v}{\kappa_w+\kappa_v}\int_{-\infty}^{+\infty}dx\left[\theta(x_{A1}-x)e^{\kappa_w(x-x_{A1})}+\theta(x-x_{A1})e^{-\kappa_v(x-x_{A1})}\right]
\\
\times\left[-\frac{\hbar^2}{2m}\frac{d^2}{dx^2}+V\sum_{l=-N}^N[\delta(x-x_{Al})+\delta(x-x_{Bl})]\right]\left[\theta(x_{A1}-x)e^{\kappa_w(x-x_{A1})}+\theta(x-x_{A1})e^{-\kappa_v(x-x_{A1})}\right]
\\
=\frac{2\kappa_w\kappa_v}{\kappa_w+\kappa_v}\bigg\{\int_{-\infty}^{+\infty}dxV\sum_{l=-N}^N[\delta(x-x_{Al})+\delta(x-x_{Bl})]\left[\theta(x_{A1}-x)e^{\kappa_w(x-x_{A1})}+\theta(x-x_{A1})e^{-\kappa_v(x-x_{A1})}\right]^2
\\
-\frac{\hbar^2}{2m}\int_{-\infty}^{+\infty}dx\left[\theta(x_{A1}-x)e^{\kappa_w(x-x_{A1})}+\theta(x-x_{A1})e^{-\kappa_v(x-x_{A1})}\right]\times
\\
\left[\kappa_w^2\theta(x_{A1}-x)e^{\kappa_w(x-x_{A1})}+\kappa_v^2\theta(x-x_{A1})e^{-\kappa_v(x-x_{A1})}\right]
\\
-\frac{\hbar^2}{2m}\int_{-\infty}^{+\infty}dx\left[\theta(x_{A1}-x)e^{\kappa_w(x-x_{A1})}+\theta(x-x_{A1})e^{-\kappa_v(x-x_{A1})}\right]
\\
\times\left[\delta'(x_{A1}-x)e^{\kappa_w(x-x_{A1})}+\delta'(x-x_{A1})e^{-\kappa_v(x-x_{A1})}-2\delta(x-x_{A1})\left(\kappa_we^{\kappa_w(x-x_{A1})}+\kappa_ve^{-\kappa_v(x-x_{A1})}\right)\right]\bigg\}
\\
=\frac{2\kappa_w\kappa_v}{\kappa_w+\kappa_v}\bigg\{\int_{-\infty}^{x_{A1}}dx\left[-\frac{\hbar^2\kappa_w^2}{2m}+V[\delta(x-x_{A0})+\delta(x-x_{B0})+\delta(x-x_{A1})]\right]e^{2\kappa_w(x-x_{A1})}
\\
+\int_{x_{A1}}^{+\infty}dx\left[-\frac{\hbar^2\kappa_v^2}{2m}+V[\delta(x-x_{A1})+\delta(x-x_{B1})+\delta(x-x_{A2})]\right]e^{-2\kappa_v(x-x_{A1})}
\\
-\frac{\hbar^2}{2m}\int_{-\infty}^{x_{A1}}dx\Big[\delta'(x_{A1}-x)e^{2\kappa_w(x-x_{A1})}+\delta'(x-x_{A1})e^{(\kappa_w-\kappa_v)(x-x_{A1})}
\\
-2\delta(x-x_{A1})\left(\kappa_we^{2\kappa_w(x-x_{A1})}+\kappa_ve^{(\kappa_w-\kappa_v)(x-x_{A1})}\right)\Big]
\\
-\frac{\hbar^2}{2m}\int_{x_{A1}}^{+\infty}dx\Big[\delta'(x_{A1}-x)e^{(\kappa_w-\kappa_v)(x-x_{A1})}+\delta'(x-x_{A1})e^{-2\kappa_v(x-x_{A1})}
\\
-2\delta(x-x_{A1})\left(\kappa_we^{(\kappa_w-\kappa_v)(x-x_{A1})}+\kappa_ve^{-2\kappa_v(x-x_{A1})}\right)\Big].
\end{multline}
It is a standard result that:
\begin{equation}
\int_a^bdxf(x)\frac{d^n}{dx^n}\delta(\pm x-c)=\pm(-1)^nf^{(n)}(c)\theta(c-a)\theta(b-c),
\end{equation}
and so:
\begin{multline}
\epsilon_{AA}=
\\
\frac{2\kappa_w\kappa_v}{\kappa_w+\kappa_v}\bigg\{\frac{E_w}{2\kappa_w}+V\left(e^{2\kappa_w(x_{A0}-x_{A1})}+e^{2\kappa_w(x_{B0}-x_{A1})}+\frac{1}{2}\right)
+\frac{E_v}{2\kappa_v}+V\left(\frac{1}{2}+e^{-2\kappa_v(x_{B1}-x_{A1})}+e^{-2\kappa_v(x_{A2}-x_{A1})}\right)
\\
-\frac{\hbar^2}{2m}\left[\kappa_w-(\kappa_w-\kappa_v)-\kappa_w-\kappa_v+(\kappa_w-\kappa_v)+\kappa_v
-\kappa_w-\kappa_v\right]\bigg\}
\\
=\frac{2\kappa_w\kappa_v}{\kappa_w+\kappa_v}\left[\frac{E_w}{2\kappa_w}+\frac{E_v}{2\kappa_v}+V\left(1+e^{-2\kappa_ww}+e^{-2\kappa_vv}+e^{-2\kappa_wd}+e^{-2\kappa_vd}\right)
+\frac{\hbar^2}{2m}(\kappa_w+\kappa_v)\right]
\\
=\frac{\kappa_w\kappa_v}{\kappa}V\left(1+e^{-2\kappa_ww}+e^{-2\kappa_vv}+e^{-2\kappa_wd}+e^{-2\kappa_vd}\right)
+\frac{\hbar^2}{2m}\kappa_w\kappa_v.
\end{multline}
Thus, the on-site potential is:
\begin{equation}
\epsilon_{AA}=\frac{1}{\kappa}\kappa_w\kappa_vV\left(1+e^{-2\kappa_ww}+e^{-2\kappa_vv}+e^{-2\kappa_wd}+e^{-2\kappa_vd}\right)
-\frac{1}{2\kappa}(\kappa_vE_w+\kappa_wE_v).
\end{equation}
Now we find the on-site potential for the B sublattice as:
\begin{multline}
\epsilon_{BB}=\frac{2\kappa_w\kappa_v}{\kappa_w+\kappa_v}\int_{-\infty}^{+\infty}dx\left[\theta(x_{B1}-x)e^{\kappa_v(x-x_{B1})}+\theta(x-x_{B1})e^{-\kappa_w(x-x_{B1})}\right]\times
\\
\left[-\frac{\hbar^2}{2m}\frac{d^2}{dx^2}+V\sum_{l=-N}^N[\delta(x-x_{Al})+\delta(x-x_{Bl})]\right]\left[\theta(x_{B1}-x)e^{\kappa_v(x-x_{B1})}+\theta(x-x_{B1})e^{-\kappa_w(x-x_{B1})}\right]
\\
=\frac{2\kappa_w\kappa_v}{\kappa_w+\kappa_v}\bigg\{\int_{-\infty}^{x_{B1}}dx\left[E_v+V[\delta(x-x_{B0})+\delta(x-x_{A1})+\delta(x-x_{B1})]\right]e^{2\kappa_v(x-x_{B1})}
\\
\int_{x_{B1}}^{+\infty}dx\left[E_w+V[\delta(x-x_{B1})+\delta(x-x_{A2})+\delta(x-x_{B2})]\right]e^{-2\kappa_w(x-x_{B1})}
\\
-\frac{\hbar^2}{2m}\int_{-\infty}^{x_{B1}}dx\Big[\delta'(x_{B1}-x)e^{2\kappa_v(x-x_{B1})}+\delta'(x-x_{B1})e^{(\kappa_v-\kappa_w)(x-x_{B1})}
\\
-2\delta(x-x_{B1})\left(\kappa_ve^{2\kappa_v(x-x_{B1})}+\kappa_we^{(\kappa_v-\kappa_w)(x-x_{B1})}\right)\Big]
\\
-\frac{\hbar^2}{2m}\int_{x_{B1}}^{+\infty}dx\Big[\delta'(x_{B1}-x)e^{(\kappa_v-\kappa_w)(x-x_{B1})}+\delta'(x-x_{B1})e^{-2\kappa_w(x-x_{B1})}
\\
-2\delta(x-x_{B1})\left(\kappa_ve^{(\kappa_v-\kappa_w)(x-x_{B1})}+\kappa_we^{-2\kappa_w(x-x_{B1})}\right)\Big],
\end{multline}
which evaluates as:
\begin{multline}
\epsilon_{BB}=
\\
\frac{2\kappa_w\kappa_v}{\kappa_w+\kappa_v}\bigg\{\frac{E_v}{2\kappa_v}+V\left(e^{2\kappa_v(x_{B0}-x_{B1})}+e^{2\kappa_v(x_{A1}-x_{B1})}+\frac{1}{2}\right)
+\frac{E_w}{2\kappa_w}+V\left(\frac{1}{2}+e^{-2\kappa_w(x_{A2}-x_{B1})}+e^{-2\kappa_w(x_{B2}-x_{B1})}\right)
\\
-\frac{\hbar^2}{2m}\left[\kappa_v-(\kappa_v-\kappa_w)-\kappa_v-\kappa_w+(\kappa_v-\kappa_w)
+\kappa_w-\kappa_v-\kappa_w\right]\bigg\}
\\
=\frac{2\kappa_w\kappa_v}{\kappa_w+\kappa_v}\left[\frac{E_w}{2\kappa_w}+\frac{E_v}{2\kappa_v}+V\left(1+e^{-2\kappa_vv}+e^{-2\kappa_ww}+e^{-2\kappa_wd}+e^{-2\kappa_vd}\right)
+\frac{\hbar^2}{2m}(\kappa_w+\kappa_v)\right]
\\
=\frac{\kappa_w\kappa_v}{\kappa}V\left(1+e^{-2\kappa_vv}+e^{-2\kappa_ww}+e^{-2\kappa_wd}+e^{-2\kappa_vd}\right)
-\frac{1}{2\kappa}(\kappa_vE_w+\kappa_wE_v),
\end{multline}
and thus we see that $\epsilon_{AA}=\epsilon_{BB}$ as it ought to be. In fact, this expression may be simplified further since $V/\kappa=-\hbar^2/m$ to become:
\begin{multline}
\epsilon_{AA}=\epsilon_{BB}=\epsilon=\frac{-\hbar^2}{m}\kappa_w\kappa_v\left(1+e^{-2\kappa_vv}+e^{-2\kappa_ww}+e^{-2\kappa_wd}+e^{-2\kappa_vd}\right)
+\frac{\hbar^2}{2m}\kappa_w\kappa_v
\\
=\frac{-\hbar^2}{2m}\kappa_w\kappa_v\left[1+2\left(e^{-2\kappa_vv}+e^{-2\kappa_ww}+e^{-2\kappa_vd}+e^{-2\kappa_wd}\right)\right]
\end{multline}

Now, moving onto the intra-cell hopping term $t_{AB}$ with $i=1$, $j=1$:
\begin{multline}
t_{AB}=e^{ik(x_{B1}-x_{A1})}\frac{2\kappa_w\kappa_v}{\kappa_w+\kappa_v}\bigg\{\int_{-\infty}^{+\infty}dx\left[\theta(x_{A1}-x)e^{\kappa_w(x-x_{A1})}+\theta(x-x_{A1})e^{-\kappa_v(x-x_{A1})}\right]\times
\\
\left[-\frac{\hbar^2}{2m}\frac{d^2}{dx^2}+V\sum_{l=-N}^N[\delta(x-x_{Al})+\delta(x-x_{Bl})]\right]\left[\theta(x_{B1}-x)e^{\kappa_v(x-x_{B1})}+\theta(x-x_{B1})e^{-\kappa_w(x-x_{B1})}\right]
\\
-\frac{\hbar^2}{2m}\int_{-\infty}^{+\infty}dx\left[\theta(x_{A1}-x)e^{\kappa_w(x-x_{A1})}+\theta(x-x_{A1})e^{-\kappa_v(x-x_{A1})}\right]
\\
\times\left[\delta'(x_{B1}-x)e^{\kappa_v(x-x_{B1})}+\delta'(x-x_{B1})e^{-\kappa_w(x-x_{B1})}-2\delta(x-x_{B1})\left(\kappa_ve^{\kappa_v(x-x_{B1})}+\kappa_we^{-\kappa_w(x-x_{B1})}\right)\right]
\\
=e^{ik(x_{B1}-x_{A1})}\frac{2\kappa_w\kappa_v}{\kappa_w+\kappa_v}\bigg\{
\int_{-\infty}^{x_{A1}}dx\left[E_v+V[\delta(x-x_{A0})+\delta(x-x_{B0})+\delta(x-x_{A1})]\right]e^{(\kappa_w+\kappa_v)x-(\kappa_vx_{B1}+\kappa_wx_{A1})}
\\
+\int_{x_{A1}}^{x_{B1}}dx\left[E_v+V[\delta(x-x_{A1})+\delta(x-x_{B1})]\right]e^{\kappa_v(x_{A1}-x_{B1})}
\\
+\int_{x_{B1}}^{+\infty}dx\left[E_w+V[\delta(x-x_{B1})+\delta(x-x_{A2})+\delta(x-x_{B2})]\right]e^{-(\kappa_w+\kappa_v)x+(\kappa_wx_{B1}+\kappa_vx_{A1})}
\\
-\frac{\hbar^2}{2m}\int_{x_{A1}}^{+\infty}dx\Big[\delta'(x_{B1}-x)e^{\kappa_v(x_{A1}-x_{B1})}+\delta'(x-x_{B1})e^{-(\kappa_v+\kappa_w)x+\kappa_vx_{A1}+\kappa_wx_{B1})}
\\
-2\delta(x-x_{B1})\left(\kappa_ve^{\kappa_v(x_{A1}-x_{B1})}
+\kappa_we^{-(\kappa_v+\kappa_w)x+\kappa_vx_{A1}+\kappa_wx_{B1}}\right)\Big]
\bigg\}.
\end{multline}
This becomes:
\begin{multline}
t_{AB}=e^{ik(x_{B1}-x_{A1})}\frac{2\kappa_w\kappa_v}{\kappa_w+\kappa_v}\bigg\{\left[\frac{E_v+E_w}{2(\kappa_w+\kappa_v)}+[E_v(x_{B1}-x_{A1})+V]\right]e^{\kappa_v(x_{A1}-x_{B1})}
\\
+V\Big(e^{(\kappa_w+\kappa_v)x_{A0}-\kappa_vx_{B1}-\kappa_wx_{A1}}
+e^{(\kappa_w+\kappa_v)x_{B0}-\kappa_vx_{B1}-\kappa_wx_{A1}}
+\frac{1}{2}e^{(\kappa_w+\kappa_v)x_{A1}-\kappa_vx_{B1}-\kappa_wx_{A1}}\Big)
\\
+V\left(\frac{1}{2}e^{-(\kappa_w+\kappa_v)x_{B1}+\kappa_wx_{B1}+\kappa_vx_{A1}}
+e^{-(\kappa_w+\kappa_v)x_{A2}+\kappa_wx_{B1}+\kappa_vx_{A1}}
+e^{-(\kappa_w+\kappa_v)x_{B2}+\kappa_wx_{B1}+\kappa_vx_{A1}}\right)
\\
-\frac{\hbar^2}{2m}\left[(\kappa_w+\kappa_v)e^{-(\kappa_v+\kappa_w)x_{B1}+\kappa_vx_{A1}+\kappa_wx_{B1})}-2\kappa_ve^{\kappa_v(x_{A1}-x_{B1})}-2\kappa_we^{-(\kappa_v+\kappa_w)x_{B1}+\kappa_vx_{A1}+\kappa_wx_{B1}}\right]\bigg\}
\\
=e^{ikv}\frac{2\kappa_w\kappa_v}{\kappa_w+\kappa_v}\bigg\{\left(\frac{E_v}{2\kappa}+\frac{E_w}{2\kappa}+E_vv+V\right)e^{-\kappa_vv}+\frac{\hbar^2}{2m}(\kappa_w+\kappa_v)e^{-\kappa_vv}
\\
+V\left(e^{-\kappa_wv-(\kappa_w+\kappa_v)d}
+e^{-\kappa_vw-\kappa_wd}+\frac{1}{2}e^{-\kappa_wv}+\frac{1}{2}e^{-\kappa_vv}+e^{-\kappa_ww-\kappa_vd}
+e^{-\kappa_vv-(\kappa_w+\kappa_v)d}\right)\bigg\}.
\end{multline}
Now, again $\kappa_v+\kappa_w=2\kappa$, and so:
\begin{multline}
t_{AB}=e^{ikv}\frac{\kappa_w\kappa_v}{\kappa}\bigg\{\left(\frac{E_v+E_w}{2\kappa}+E_vv+V+\frac{\hbar^2\kappa}{m}\right)e^{-\kappa_vv}+\frac{V}{2}\left(e^{-\kappa_vv}+e^{-\kappa_wv}\right)
\\
+V\left(e^{-\kappa_vw-\kappa_wd}+e^{-\kappa_ww-\kappa_vd}\right)+V\left(e^{-\kappa_vv}
+e^{-\kappa_wv}\right)e^{-2\kappa d}\bigg\}.
\end{multline}
Thus, to first order (ignoring any $e^{-\kappa d}$ terms) and recalling that $\kappa=-mV/\hbar^2$:
\begin{equation}
t_{AB}=\frac{\kappa_w\kappa_v}{\kappa}\left[\left(\frac{E_v+E_w}{2\kappa}+E_vv\right)e^{-\kappa_vv}+\frac{V}{2}\left(e^{-\kappa_vv}+e^{-\kappa_wv}\right)\right]e^{ikv}.
\end{equation}

Now, we again move onto the inter unit cell hopping $\bar{t}_{BA}$ between $i=1$ and $j=2$, which is given by:
\begin{multline}
\bar{t}_{BA}=e^{ik(x_{A2}-x_{B1})}\frac{2\kappa_w\kappa_v}{\kappa_w+\kappa_v}\bigg\{\int_{-\infty}^{+\infty}dx\left[\theta(x_{B1}-x)e^{\kappa_v(x-x_{B1})}+\theta(x-x_{B1})e^{-\kappa_w(x-x_{B1})}\right]\times
\\
\left[-\frac{\hbar^2}{2m}\frac{d^2}{dx^2}+V\sum_{l=-N}^N[\delta(x-x_{Al})+\delta(x-x_{Bl})]\right]\left[\theta(x_{A2}-x)e^{\kappa_w(x-x_{A2})}+\theta(x-x_{A2})e^{-\kappa_v(x-x_{A2})}\right]
\\
-\frac{\hbar^2}{2m}\int_{-\infty}^{+\infty}dx\left[\theta(x_{B1}-x)e^{\kappa_v(x-x_{B1})}+\theta(x-x_{B1})e^{-\kappa_w(x-x_{B1})}\right]
\\
\times\left[\delta'(x_{A2}-x)e^{\kappa_w(x-x_{A2})}+\delta'(x-x_{A2})e^{-\kappa_v(x-x_{A2})}-2\delta(x-x_{A2})\left(\kappa_we^{\kappa_w(x-x_{A2})}+\kappa_ve^{-\kappa_v(x-x_{A2})}\right)\right]
\\
=e^{ik(x_{A2}-x_{B1})}\frac{2\kappa_w\kappa_v}{\kappa_w+\kappa_v}\bigg\{
\int_{-\infty}^{x_{B1}}dx\left[E_w+V[\delta(x-x_{B0})+\delta(x-x_{A1})+\delta(x-x_{B1})]\right]e^{(\kappa_w+\kappa_v)x-(\kappa_wx_{A2}+\kappa_vx_{B1})}
\\
+\int_{x_{B1}}^{x_{A2}}dx\left[E_w+V[\delta(x-x_{B1})+\delta(x-x_{A2})]\right]e^{\kappa_w(x_{B1}-x_{A2})}
\\
+\int_{x_{A2}}^{+\infty}dx\left[E_v+V[\delta(x-x_{A2})+\delta(x-x_{B2})+\delta(x-x_{A3})]\right]e^{-(\kappa_w+\kappa_v)x+(\kappa_vx_{A2}+\kappa_wx_{B1})}
\\
-\frac{\hbar^2}{2m}\int_{x_{B1}}^{+\infty}dx\Big[\delta'(x_{A2}-x)e^{\kappa_w(x_{B1}-x_{A2})}+\delta'(x-x_{A2})e^{-(\kappa_v+\kappa_w)x+(\kappa_vx_{A2}+\kappa_wx_{B1})}
\\
-2\delta(x-x_{A2})\left(\kappa_we^{\kappa_w(x_{B1}-x_{A2})}
+\kappa_ve^{-(\kappa_v+\kappa_w)x+\kappa_wx_{B1}+\kappa_vx_{A2}}\right)\Big].
\bigg\}
\end{multline}
This becomes:
\begin{multline}
\bar{t}_{BA}=e^{ik(x_{A2}-x_{B1})}\frac{2\kappa_w\kappa_v}{\kappa_w+\kappa_v}\bigg\{\left[\frac{E_v+E_w}{2(\kappa_w+\kappa_v)}+E_w(x_{A2}-x_{B1})+V\right]e^{\kappa_w(x_{B1}-x_{A2})}
\\
+V\left(e^{(\kappa_w+\kappa_v)x_{B0}-(\kappa_wx_{A2}+\kappa_vx_{B1})}
+e^{(\kappa_w+\kappa_v)x_{A1}-(\kappa_wx_{A2}+\kappa_vx_{B1})}
+\frac{1}{2}e^{(\kappa_w+\kappa_v)x_{B1}-(\kappa_wx_{A2}+\kappa_vx_{B1})}\right)
\\
+V\left(\frac{1}{2}e^{-(\kappa_w+\kappa_v)x_{A2}+(\kappa_vx_{A2}+\kappa_wx_{B1})}
+e^{-(\kappa_w+\kappa_v)x_{B2}+(\kappa_vx_{A2}+\kappa_wx_{B1})}
+e^{-(\kappa_w+\kappa_v)x_{A3}+(\kappa_vx_{A2}+\kappa_wx_{B1})}\right)
\\
-\frac{\hbar^2}{2m}\left[(\kappa_w+\kappa_v)e^{-(\kappa_v+\kappa_w)x_{A2}+\kappa_vx_{A2}
+\kappa_wx_{B1}}-2\kappa_we^{\kappa_w(x_{B1}-x_{A2})}-2\kappa_ve^{-(\kappa_v+\kappa_w)x_{A2}+\kappa_wx_{B1}+\kappa_vx_{A2})}\right]
\\
=e^{ikw}\frac{\kappa_w\kappa_v}{\kappa}\bigg\{\left[\frac{E_v+E_w}{4\kappa}+E_ww+V\right]e^{-\kappa_ww}+\frac{\hbar^2}{2m}(\kappa_w+\kappa_v)e^{-\kappa_ww}
\\
+V\left(e^{-\kappa_ww-(\kappa_v+\kappa_w)d}+e^{-\kappa_vv-\kappa_wd}+\frac{1}{2}e^{-\kappa_ww}+\frac{1}{2}e^{-\kappa_vw}
+e^{-\kappa_wv-\kappa_wd}+e^{-\kappa_vw-(\kappa_v+\kappa_w)d}\right)\bigg\}
\\
=e^{ikw}\frac{\kappa_w\kappa_v}{\kappa}\bigg\{\left(\frac{E_v+E_w}{2\kappa}+E_ww+V+\frac{\hbar^2\kappa}{m}\right)e^{-\kappa_ww}+\frac{V}{2}\left(e^{-\kappa_vw}+e^{-\kappa_ww}\right)
\\
+V\left(e^{-\kappa_ww}+e^{-\kappa_vw}\right)e^{-2\kappa d}+V\left(e^{-\kappa_vv}+e^{-\kappa_wv}\right)e^{-\kappa_wd}\bigg\}.
\end{multline}
This may too be seen to reduce to the previous result. So, to first order:
\begin{equation}
\bar{t}_{BA}=\frac{\kappa_w\kappa_v}{\kappa}\left[\left(\frac{E_v+E_w}{2\kappa}+E_ww\right)e^{-\kappa_ww}+\frac{V}{2}\left(e^{-\kappa_vw}+e^{-\kappa_ww}\right)\right]e^{ikw}.
\end{equation}

Thus, to first order in the nearest-neighbour interactions, the effective tight-binding 2x2 matrix Hamiltonian is:
\begin{equation}
H=
\begin{pmatrix}
\epsilon&t_{AB}+\bar{t}_{AB}
\\
t_{BA}+\bar{t}_{BA}&\epsilon
\end{pmatrix},
\end{equation}
where $t_{AB}=t_{BA}^*$, $\bar{t}_{AB}=\bar{t}_{BA}^*$, and:
\begin{align}
\epsilon&=\frac{-\hbar^2}{2m}\kappa_w\kappa_v\left[1+2\left(e^{-2\kappa_vv}+e^{-2\kappa_ww}+e^{-2\kappa_vd}+e^{-2\kappa_wd}\right)\right],
\\
t_{AB}&=\frac{\kappa_w\kappa_v}{\kappa}\left[\left(\frac{1}{2\kappa}(E_v+E_w)+E_vv\right)e^{-\kappa_vv}+\frac{V}{2}\left(e^{-\kappa_vv}+e^{-\kappa_wv}\right)\right]e^{ikv}.
\\
\bar{t}_{AB}&=\frac{\kappa_w\kappa_v}{\kappa}\left[\left(\frac{1}{2\kappa}(E_v+E_w)+E_ww\right)e^{-\kappa_ww}+\frac{V}{2}\left(e^{-\kappa_vw}+e^{-\kappa_ww}\right)\right]e^{-ikw}.
\end{align}

On the other hand, the matrix elements of the overlap matrix may be found simply as follows. The on-diagonals are equal to one because the basis wavefunctions are normalised correctly. Then the off-diagonals follow as:
\begin{multline}
\eta_{AB}=\frac{2\kappa_w\kappa_v}{\kappa_w+\kappa_v}e^{ik(x_{B1}-x_{A1})}
\\
\times\int_{-\infty}^{+\infty}dx\left[\theta(x_{A1}-x)e^{\kappa_w(x-x_{A1})}+\theta(x-x_{A1})e^{-\kappa_v(x-x_{A1})}\right]\left[\theta(x_{B1}-x)e^{\kappa_v(x-x_{B1})}+\theta(x-x_{B1})e^{-\kappa_w(x-x_{B1})}\right]
\\
=\frac{\kappa_w\kappa_v}{\kappa}e^{ik(x_{B1}-x_{A1})}
\times\left[\int_{-\infty}^{x_{A1}}dxe^{(\kappa_w+\kappa_v)x
-\kappa_wx_{A1}-\kappa_vx_{B1}}
+\int_{x_{A1}}^{x_{B1}}dxe^{\kappa_v(x_{A1}-x_{B1})}\right.
\\
\left.+\int_{x_{A1}}^{+\infty}dxe^{-(\kappa_w+\kappa_v)x
+\kappa_vx_{A1}+\kappa_wx_{B1}}\right]=\frac{\kappa_w\kappa_v}{\kappa}\left[\frac{1}{2\kappa}\left(e^{-\kappa_vv}+e^{-\kappa_wv}\right)+ve^{-\kappa_vv}\right]e^{ikv},
\end{multline}
and:
\begin{multline}
\bar{\eta}_{BA}=\frac{2\kappa_w\kappa_v}{\kappa_w+\kappa_v}e^{ik(x_{A2}-x_{B1})}
\\
\times\int_{-\infty}^{+\infty}dx\left[\theta(x_{B1}-x)e^{\kappa_v(x-x_{B1})}+\theta(x-x_{B1})e^{-\kappa_w(x-x_{B1})}\right]\left[\theta(x_{A2}-x)e^{\kappa_w(x-x_{A2})}+\theta(x-x_{A2})e^{-\kappa_v(x-x_{A2})}\right]
\\
=\frac{\kappa_w\kappa_v}{\kappa}e^{ik(x_{A2}-x_{B1})}\left[\int_{-\infty}^{x_{B1}}dxe^{(\kappa_w+\kappa_v)x
-\kappa_vx_{B1}-\kappa_wx_{A2}}
+\int_{x_{B1}}^{x_{A2}}dxe^{\kappa_w(x_{B1}-x_{A2})}\right.
\\
\left.+\int_{x_{A2}}^{+\infty}dxe^{-(\kappa_w+\kappa_v)x
+\kappa_wx_{B1}+\kappa_vx_{A2}}\right]
=\frac{\kappa_w\kappa_v}{\kappa}\left[\frac{1}{2\kappa}\left(e^{-\kappa_vw}+e^{-\kappa_ww}\right)+we^{-\kappa_ww}\right]e^{ikw},
\end{multline}
thus:
\begin{equation}
S=
\begin{pmatrix}
1&\eta_{AB}+\bar{\eta}_{AB}
\\
\eta_{BA}+\bar{\eta}_{BA}&1
\end{pmatrix},
\end{equation}
where $\eta_{AB}=\eta^*_{B1}$ and $\bar{\eta}_{AB}=\bar{\eta}^*_{BA}$ as:
\begin{align}
\eta_{AB}&=\frac{\kappa_w\kappa_v}{\kappa}\left[\frac{1}{2\kappa}\left(e^{-\kappa_vv}+e^{-\kappa_wv}\right)+ve^{-\kappa_vv}\right]e^{ikv},
\\
\bar{\eta}_{AB}&=\frac{\kappa_w\kappa_v}{\kappa}\left[\frac{1}{2\kappa}\left(e^{-\kappa_vw}+e^{-\kappa_ww}\right)+we^{-\kappa_ww}\right]e^{-ikw}.
\end{align}
Therefore the matrices read:
\begin{equation}
H=
\begin{pmatrix}
\epsilon&h(k)
\\
h^*(k)&\epsilon
\end{pmatrix},\quad S=
\begin{pmatrix}
1&g(k)
\\
g^*(k)&1
\end{pmatrix},
\end{equation}
where $h(k)=te^{ikv}+t'e^{-ikw}$ and $g(k)=\eta e^{ikv}+\eta' e^{-ikw}$ with parameters as in Eq.~(\ref{eqn:hopparams}).


However, an important point must be made that may be overlooked about the matrix elements $h(k)$ and $g(k)$. Since they exist upon the off-diagonals of $H(k)$ and $S(k)$, they are in fact defined up to a phase. This `gauge' ambiguity is present in all tight-binding models be they the SSH model\cite{Asboth:2016} or that for graphene\cite{Bena:2009}. This ambiguity reflects in the {\it non-Hermitian} Zak phase
\begin{equation} \label{eq:non_herm_zak}
\theta_{\cal Z}=\frac{i}{2}\int_{-\pi/d}^{+\pi/d}dk\left[\langle\psi_{\rm L}|
\partial_k\psi_{\rm L}\rangle+\langle\psi_{\rm R}|
\partial_k\psi_{\rm R}\rangle\right],
\end{equation}
which would be quantised into units of $\pi$ plus the intra-unit-cell width $v$\cite{Zak:1989}. Note that, in the presence of full Hermicity, left and right eigenvectors are identical and this expression reduces to that of Eq.~(\ref{eq:herm_zak}).

Furthermore, the winding of $f(k)$ would not be well-defined as zero or one in the trivial and non-trivial regions, respectively. In Eq.~(\ref{eq:non_herm_zak}), ${\rm L},{\rm R}$ signify the normalised left and right eigenvectors of ${\cal H}(k)$ which, in general, do not coincide in non-Hermitian systems. This ambiguity is solved~\citep{Zak:1989} by making the wavefunction centro-symmetric about the unit-cell mid-point via a unitary transformation, so that no extra contribution to $\theta_{\cal Z}$ is acquired as a result of the imbalance in `polarisation' across the unit-cell. This is in fact exactly the same ambiguity encountered when dealing with the exact system\cite{Smith:2019}, which was solved Ref.~\onlinecite{Zak:1989} originally. In both cases, the objective is to make the unit-cell wavefunction centro-symmetric about the unit-cell mid-point.

Here, as is clear in the expressions for $h(k)=te^{ikv}+t'e^{-ikw}$ and $g(k)=\eta e^{ikv}+\eta'e^{-ikw}$, a similar effect is manifest in that there is a phase difference of $e^{ikv}$ between the two sites that constitute the unit-cell basis. As a result, a calculation of $\langle\psi_i|
\partial_k{\psi_j}\rangle$ will yield this phase in addition to the standard curvature contribution. Then we would have that $\theta_{\cal Z}=v,~\pi+v$ within the trivial and non-trivial regions respectively; an ill-defined and unquantised number.

Thus, instead of $h(k)= te^{ikv}+t'e^{-ikw}$ and $g(k)= \eta e^{ikv}+\eta'e^{-ikw}$ we should have $h(k)= t+t'e^{-ikd}$ and $g(k)= \eta+\eta'e^{-ikd}$. This may be accomplished simply by defining that $\tilde{H}(k)=U(k)H(k)U^{-1}(k)$, $\tilde{S}(k)=U(k)H(k)U^{-1}(k)$, and $\tilde{\psi}_k=U(k)\psi_k$ where:
\begin{equation}
U(k)=
\begin{pmatrix}
e^{ikv/2}&0
\\
0&e^{-ikv/2}
\end{pmatrix}.
\end{equation}
Then, dropping the tildes for brevity and clarity, the Hamiltonian and overlap matrices are in terms of the standard elements $h(k)=t+t'e^{-ikd}$ and $g(k)=\eta+\eta'e^{-ikd}$. Note that this transformation is one that is unitary and thus does nothing to mix the elements of the eigenvector $\bar{\psi}_k$. Its only effect is to modify the phases so as to well-define the observable quantum numbers, which are the topological invariants.

\section{Varying Baseline Potentials ($V_v,V_w$) as Opposed to the Separations}
\label{app:baseline}


For the latter case as described within the prose, we take the distances to be constant as $v=w=d/2$ with $d=1$ and vary the baseline potentials symmetrically as $V_w=+T$ and $V_v=-T$ whilst maintaining the Dirac-delta potentials strengths as $V=-10$. As a result, the wavevectors read:
\begin{equation}
\kappa_v=\kappa+\frac{T}{V},\quad\kappa_w=\kappa-\frac{T}{V}.
\end{equation}
Now, provided that $|T|$ never exceeds $V^2$, these wavevectors will always be positive as required for bound solutions. There thus comes a point in $|T|$ whereat the present theory breaks down. Furthermore, for sufficiently large $|T|$, next-nearest-neighbour interactions will begin to become influential.

In this case, the tight-binding parameters read:


\begin{align}
&\epsilon=-\frac{\hbar^2}{2m}\left(\kappa^2-\frac{T^2}{V^2}\right)\left[1+4\left(\cosh\left(\frac{Td}{V}\right)e^{-\kappa d}+\cosh\left(\frac{2Td}{V}\right)e^{-2\kappa d}\right)\right],\label{eqn:pep}
\\
&t=-\frac{\hbar^2}{2m}\left(\kappa^2-\frac{T^2}{V^2}\right)\left[\frac{1}{2\kappa}\left(\frac{1}{\kappa}\left(\kappa^2+\frac{T^2}{V^2}\right)+\left(\kappa^2+2\kappa\frac{T}{V}+\frac{T^2}{V^2}\right)d\right)e^{-Td/(2V)}
+2\cosh\left(\frac{Td}{2V}\right)\right]e^{-\kappa d/2},
\\
&t'=-\frac{\hbar^2}{2m}\left(\kappa^2-\frac{T^2}{V^2}\right)\left[\frac{1}{2\kappa}\left(\frac{1}{\kappa}\left(\kappa^2+\frac{T^2}{V^2}\right)+\left(\kappa^2-2\kappa \frac{T}{V}+\frac{T^2}{V^2}\right)d\right)e^{+Td/(2V)}
+2\cosh\left(\frac{Td}{2V}\right)\right]e^{-\kappa d/2},
\\
&\eta=\left(1-\frac{T^2}{\kappa^2V^2}\right)\left[\frac{\kappa d}{2}e^{-T d/(2V)}+\cosh\left(\frac{Td}{2V}\right)\right]e^{-\kappa d/2},
~
\eta'=\left(1-\frac{T^2}{\kappa^2V^2}\right)\left[\frac{\kappa d}{2}e^{T d/(2V)}+\cosh\left(\frac{Td}{2V}\right)\right]e^{-\kappa d/2}.\label{eqn:petap}
\end{align}

\begin{figure}
\centering
\begin{minipage}{.25\linewidth}
\begin{overpic}[width=\linewidth]{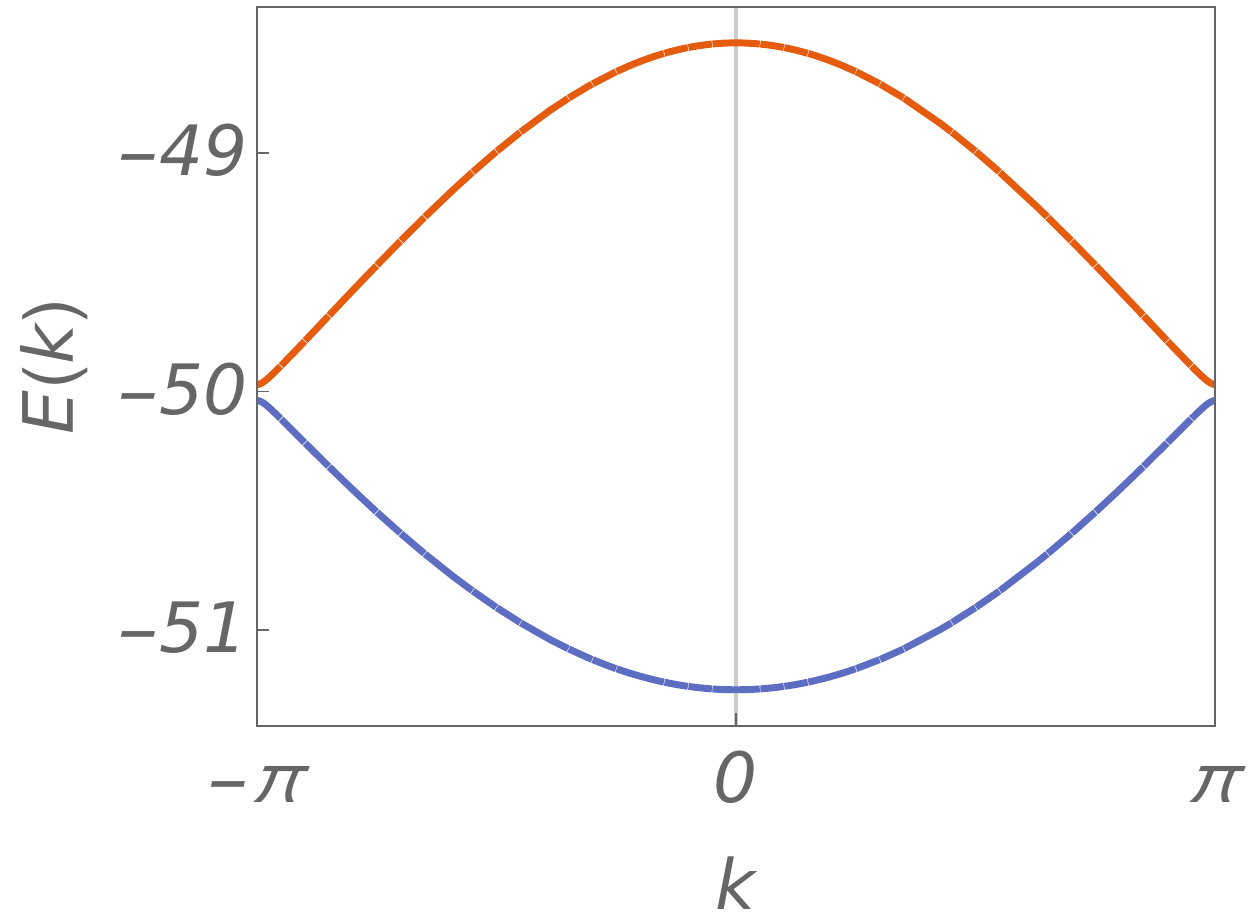}\put(22,21){(a)}
\end{overpic}
\begin{overpic}[width=\linewidth]{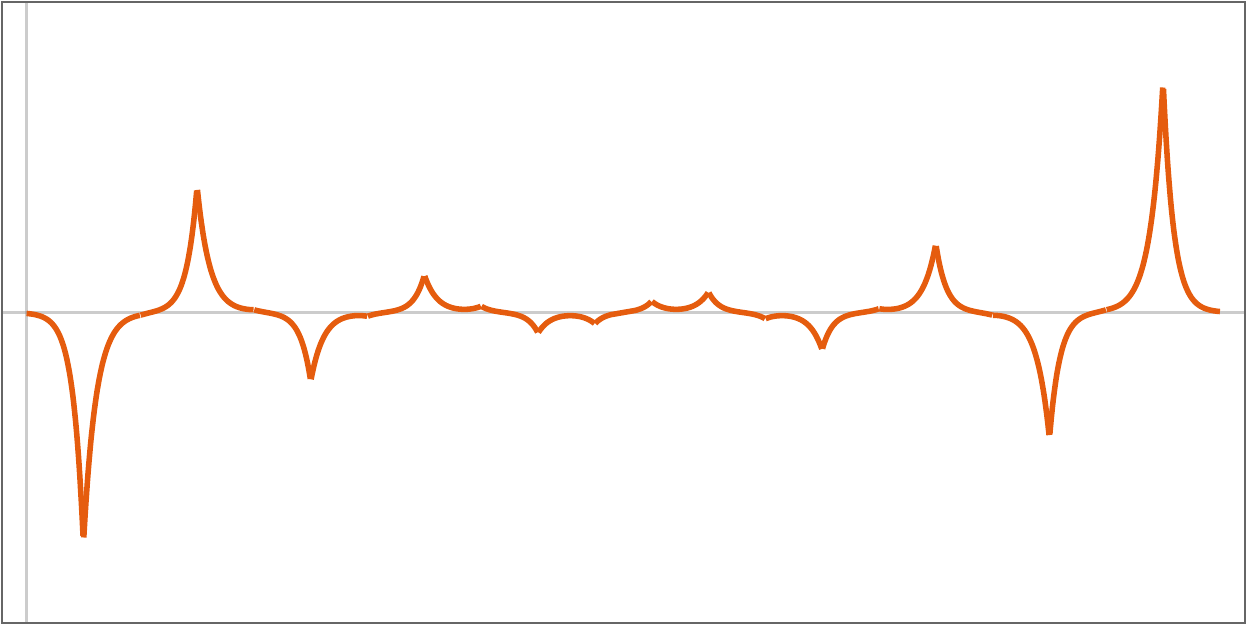}\put(89,5){(e)}
\end{overpic}
\end{minipage}%
\begin{minipage}{.25\linewidth}
\begin{overpic}[width=\linewidth]{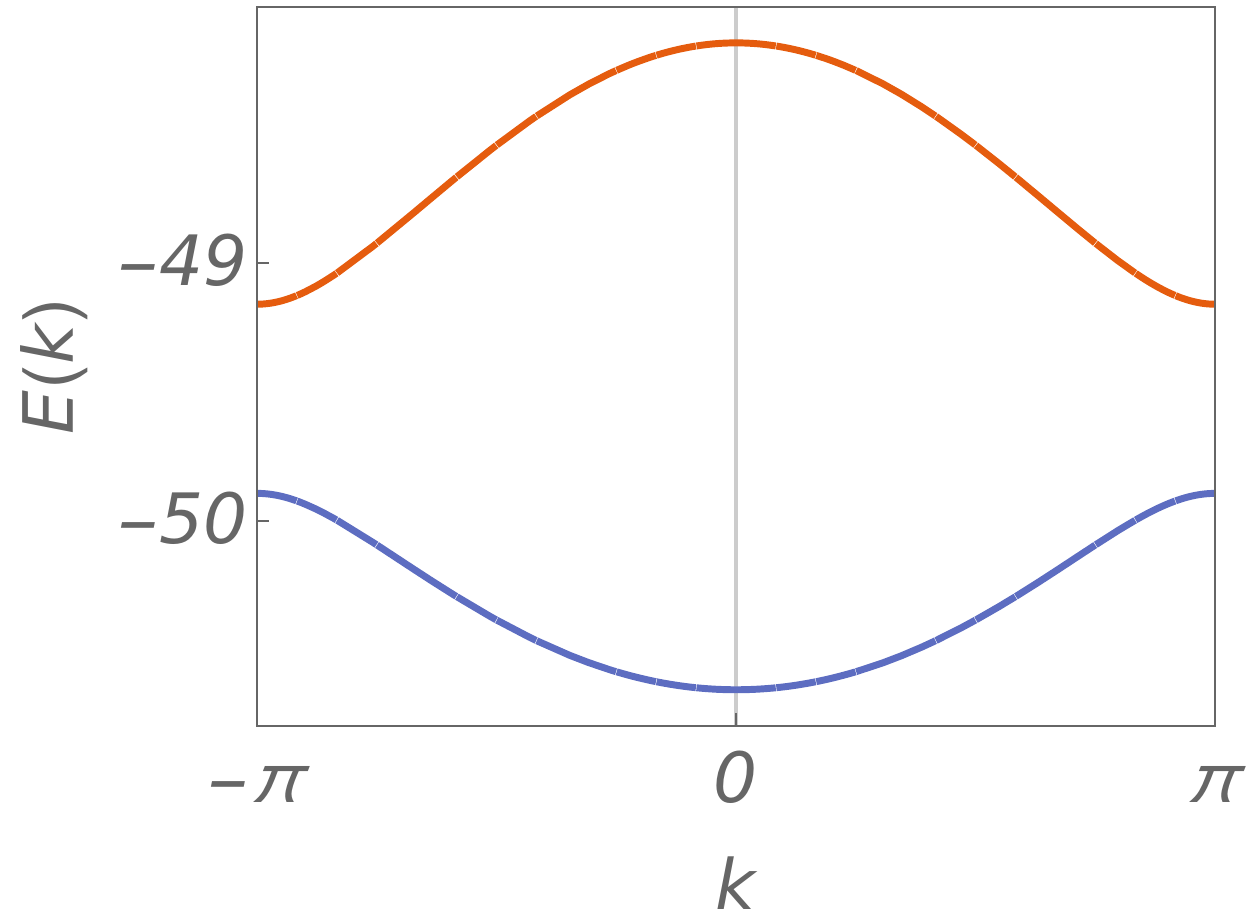}\put(22,21){(b)}
\end{overpic}
\begin{overpic}[width=\linewidth]{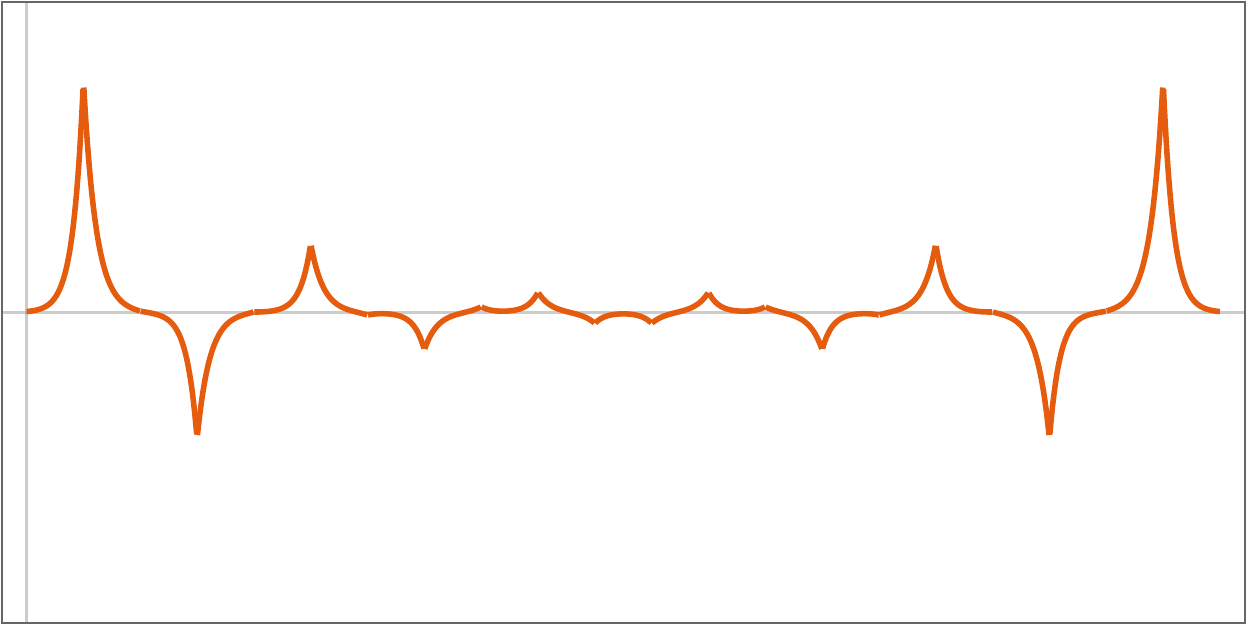}\put(3,5){(f)}
\end{overpic}
\end{minipage}%
\begin{minipage}{.25\linewidth}
\begin{overpic}[width=\linewidth]{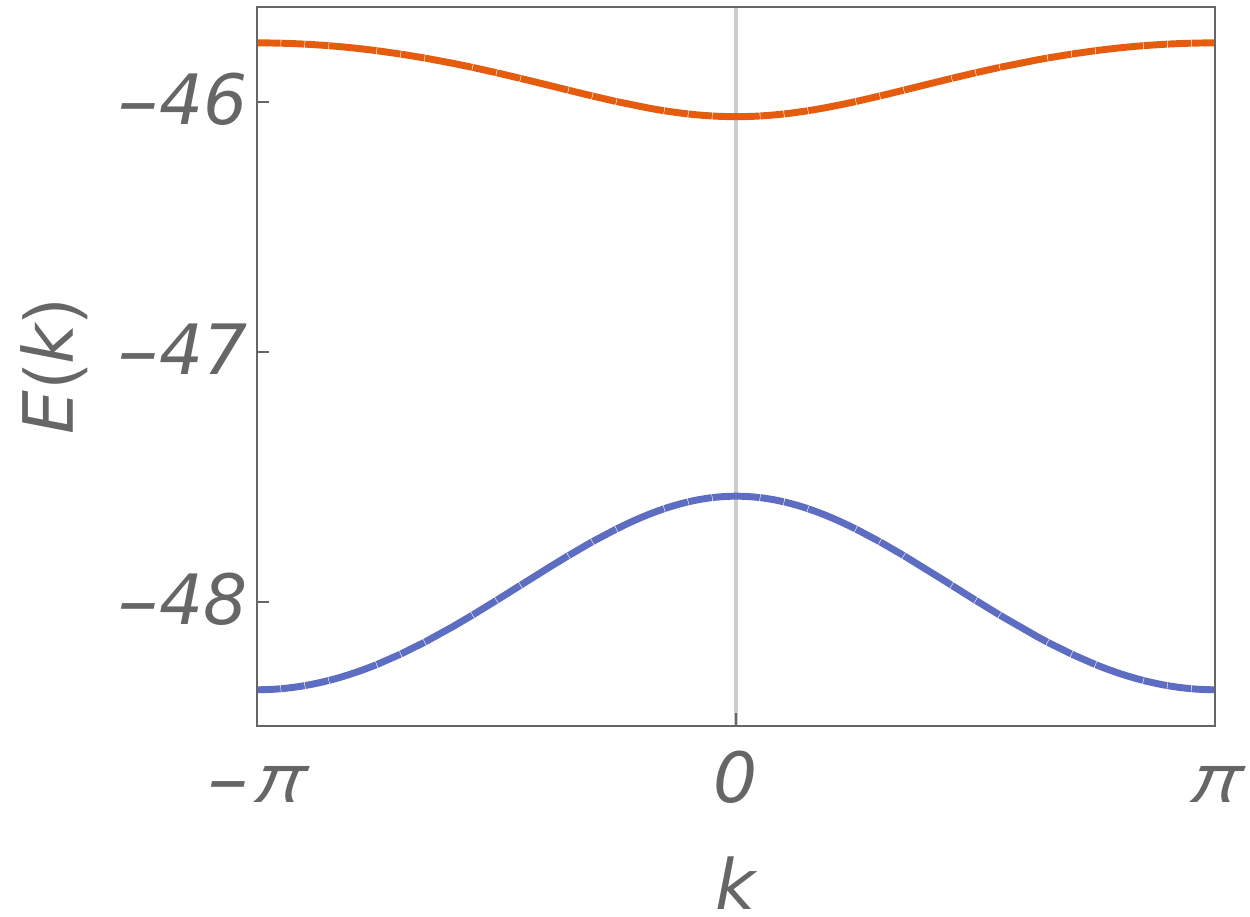}\put(22,23){(c)}
\end{overpic}
\begin{overpic}[width=\linewidth]{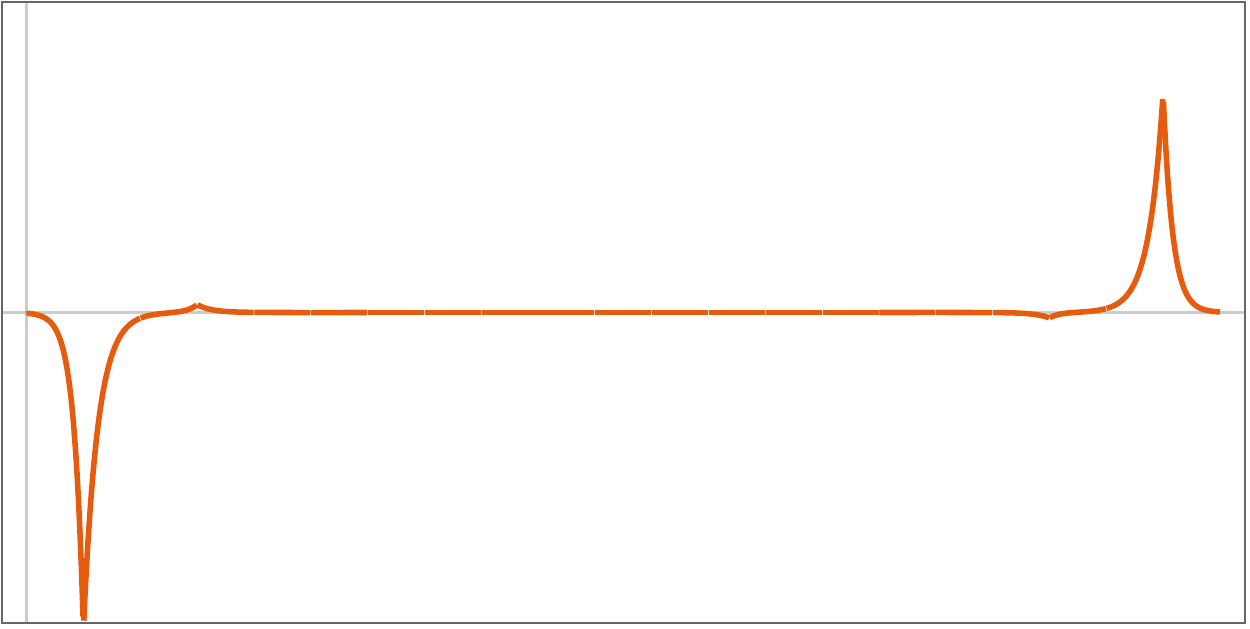}\put(89,5){(g)}
\end{overpic}
\end{minipage}%
\begin{minipage}{.25\linewidth}
\begin{overpic}[width=\linewidth]{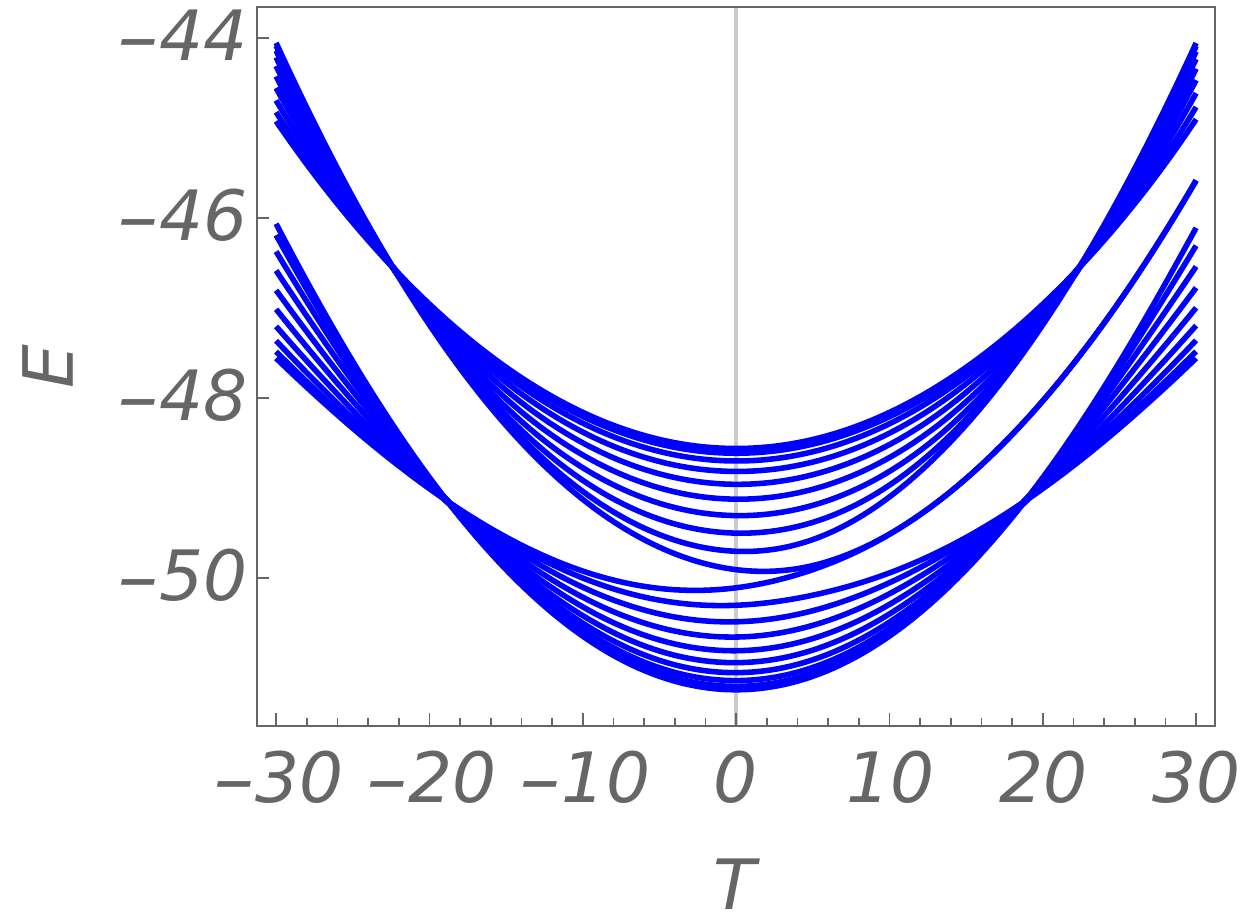}\put(22,23){(d)}
\end{overpic}
\begin{overpic}[width=\linewidth]{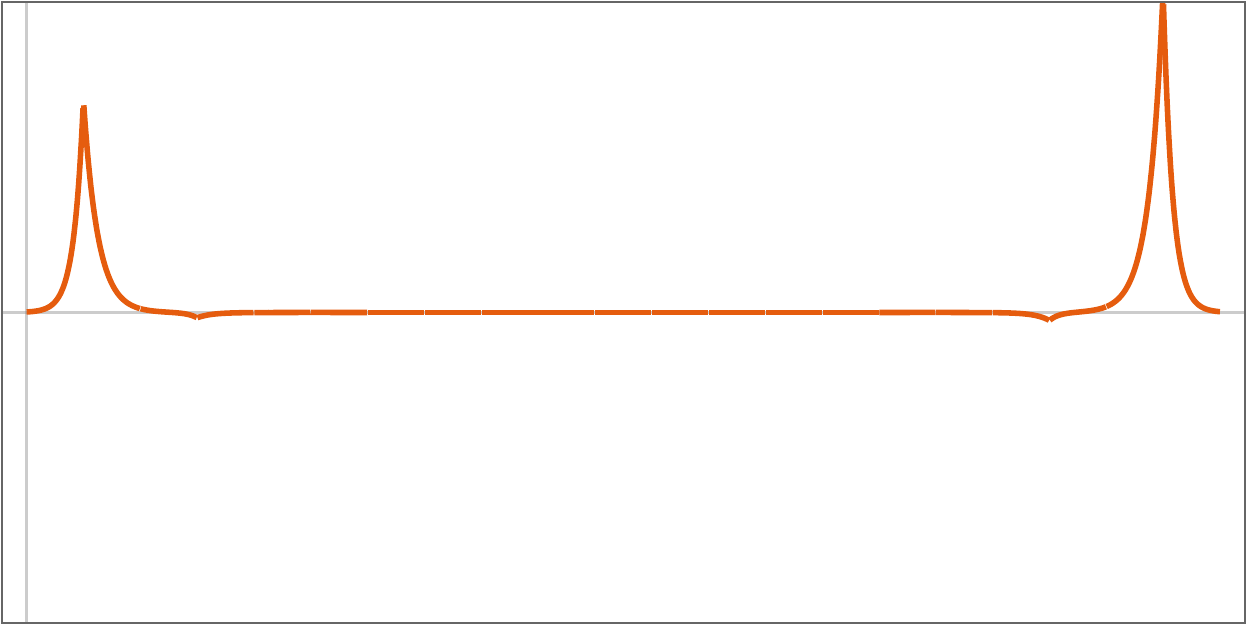}\put(3,5){(h)}
\end{overpic}
\end{minipage}
\caption{(Colour on-line) A selection of figures applying to the system wherein only the baseline potentials between the Dirac-deltas are varied. Panels (a,b,c): the bulk bands with $V_v=-1$, $V_w=1$ for (a) and with $V_v=-10$, $V_w=10$ for (b) and with $V_v=-25$, $V_w=25$ for (c). Panel (d): the finite bands with $V_v=-T$, $V_w=+T$ and boundary conditions as in (\ref{eqn:TBBC}). Panels (e,f): the 10th (e) and 11th (f) edge state wavefunctions for $T=10$. Panels (g,h): the same edge state wavefunctions for $T=20$.}
\label{fig:second}
\end{figure}

Fig.~\ref{fig:second} shows all the relevant plots of the bulk bands, finite bands and edge states of this second system under consideration. In panels (a,b), the bulk bands appear SSH-like however in panel (c), whereat $T=25$, both bands have clearly `inverted'. This phenomenon is mirrored in the finite system where the lower (upper) bulk bands touch at around $T\sim\pm20$~$(\pm25)$.

These are unphysical effects that emerge as a result of the negligence of next-nearest-neighbour hoppings. Indeed, when $T=25$ then $\kappa_w=7.5$ and $\kappa_v=12.5$ at which point the motivation for the ignorance of next-nearest-neighbour hoppings, which are proportional to $e^{TdV^{-1}}$, becomes unfounded. Such terms are non-illuminating and so will not be presented. Suffice it to say that the protection of the edge states, in this case with $V=-10$ and $d=1$, only extends up to $|T|\sim25$. After this point nothing concrete may be said about the natures of the edge states since the nearest-neighbour assumption becomes invalid.

However, within the nearest-neighbour limit of $|T|<25$, the edge states, as shown in panels (e,f,g,h), exhibit the same behaviours of being initially Hermitian, wherein the edge states are shared equally between the ends, but later non-Hermitian, wherein the edge states are attenuated/amplified at either end. This is again as a result of the absence of the $i\gamma(k)\sigma_z$ term within the bulk for $\kappa_v\sim\kappa_w$, which subsequently gains weight as $\kappa_v$ and $\kappa_w$ diverge in value.

If the Dirac-delta strengths were made more negative then the point at which next-nearest-neighbour hoppings become influential is made larger in $|T|$. However this would cause the additional effect of narrowing of the energy bands and of making the overall $T$-dependent energy shift, which affects all the bands, stronger.

\section{Numerical Proof that $F(\phi_{\rm min})$ is Never Negative}\label{app:C}

By using Eq.~(\ref{eqn:bulkparams}), we may observe that:
\begin{equation}\label{eqn:bulkbands}
E_\pm(k)=\varepsilon(k)
\pm\left(1-|g(k)|^2\right)^{-1}\sqrt{\nu^2+\omega^2+2\nu\omega\cos(kd)-\beta^2\sin^2(kd)},
\end{equation}
where $\nu=t-\eta\epsilon$, $\omega=t'-\eta'\epsilon$ and $\beta=t\eta'-t'\eta$. 

Since the band gaps are defined at the edge of the Brillouin zone at $k=\pm\pi/d$, we see that the topological phase transition is governed solely by the behaviours of $\nu$ and $\omega$ with respect to each other. The value of $\beta$ then determines whether any region {\it within} the Brillouin zone possesses complex energies and as a result introduces the concept of exceptional points.

The argument of the square-root in Eq.~(\ref{eqn:bulkbands}) may be manipulated to become:
\begin{equation}
F(\phi)=(\nu+\omega)^2+4\cos^2(\phi)\left[\nu\omega-\beta^2\sin^2(\phi)\right],
\end{equation}
where $\phi=kd/2$. This expression is clearly always positive if $\nu\omega>\beta^2$ however, when $\nu\omega<\beta^2$, it may become negative depending on the value of $\phi$. The minimum of this may be shown to occur when $\cos(2\phi_{\rm min})=-\nu\omega/\beta^2$. Thus, at this minimum point, the argument evaluates as:
\begin{equation}
F(\phi_{\rm min})=-(\omega^2-\beta^2)(\nu^2-\beta^2)\beta^{-2}.
\end{equation}
In order that $F(\phi_{\rm min})<0$ it must be that either $|\omega|>|\beta|$ and $|\nu|>|\beta|$ or $|\omega|<|\beta|$ and $|\nu|<|\beta|$. However, since it is also required that $\omega\nu<\beta^2$, the only way in which $F(\phi_{\rm min})$ may be negative is if $|\omega|<|\beta|$ and $|\nu|<|\beta|$, as was first presented in Ref.~\onlinecite{Yuce2:2018}.

This may be understood either numerically for the present system, as shown later, or qualitatively by considering the dimerised and non-bipartite limits of the system. When totally dimerised it will be that $t\to t_{\rm max}=\tau$, $t'\to0$, $\eta\to1$, $\eta'\to0$ such that $t\eta'$ and $t'\eta$ both remain non-zero so that $\beta\neq0$. Then $\nu\to\tau-\epsilon$, $\omega\to0$, $\beta\sim0$ (or in the opposite dimerised limit $\nu\to0$, $\omega\to\tau-\epsilon$, $\beta\sim0$). So in this limit, $|\omega|<|\beta|$ yet $|\nu|>|\beta|$ and so $F(\phi_{\rm min})$ cannot be negative at these limits.

\begin{figure}
\centering
\begin{minipage}{.25\linewidth}
\begin{overpic}[width=\linewidth]{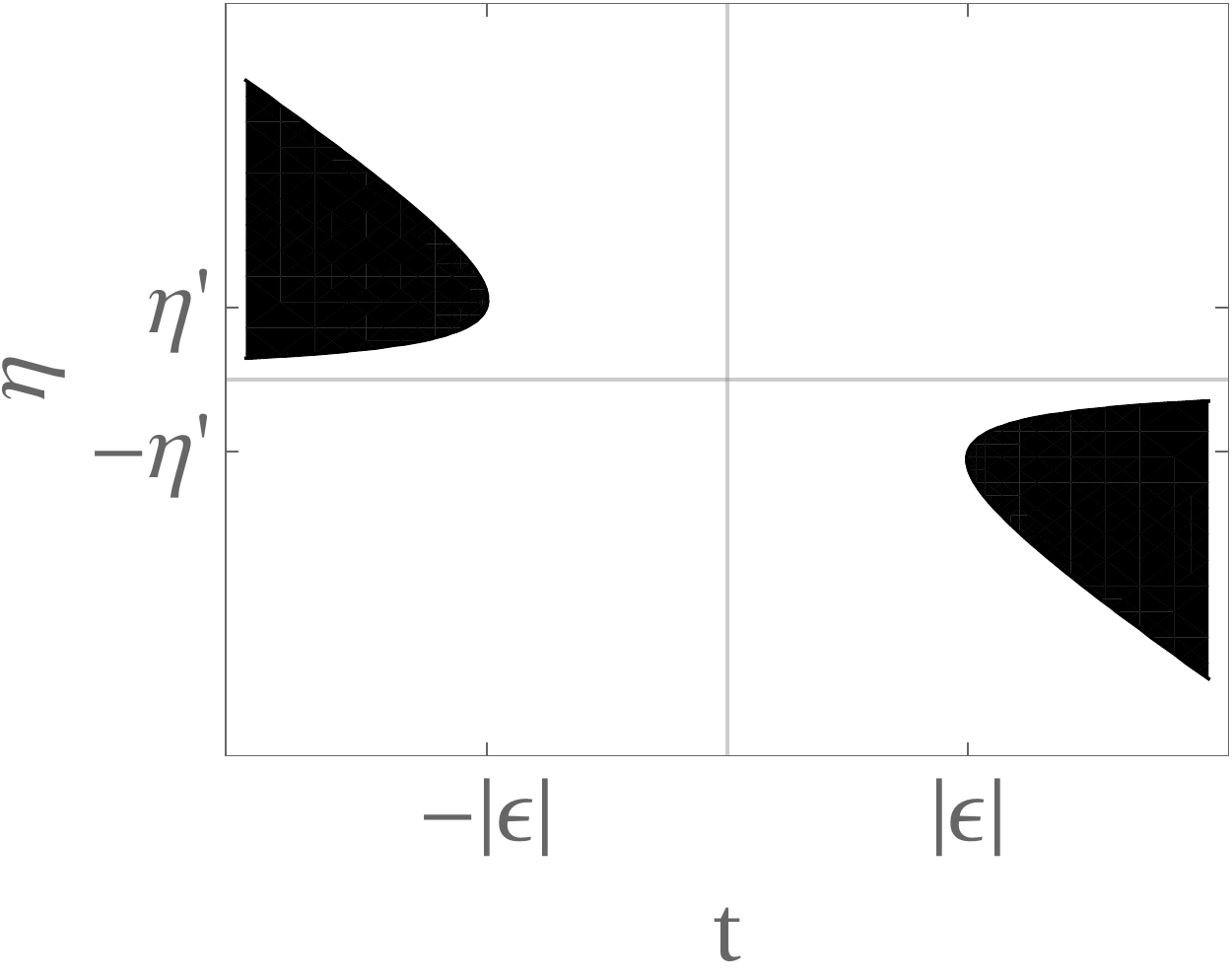}\put(88,72){(a)}
\end{overpic}
\end{minipage}%
\begin{minipage}{.25\linewidth}
\begin{overpic}[width=\linewidth]{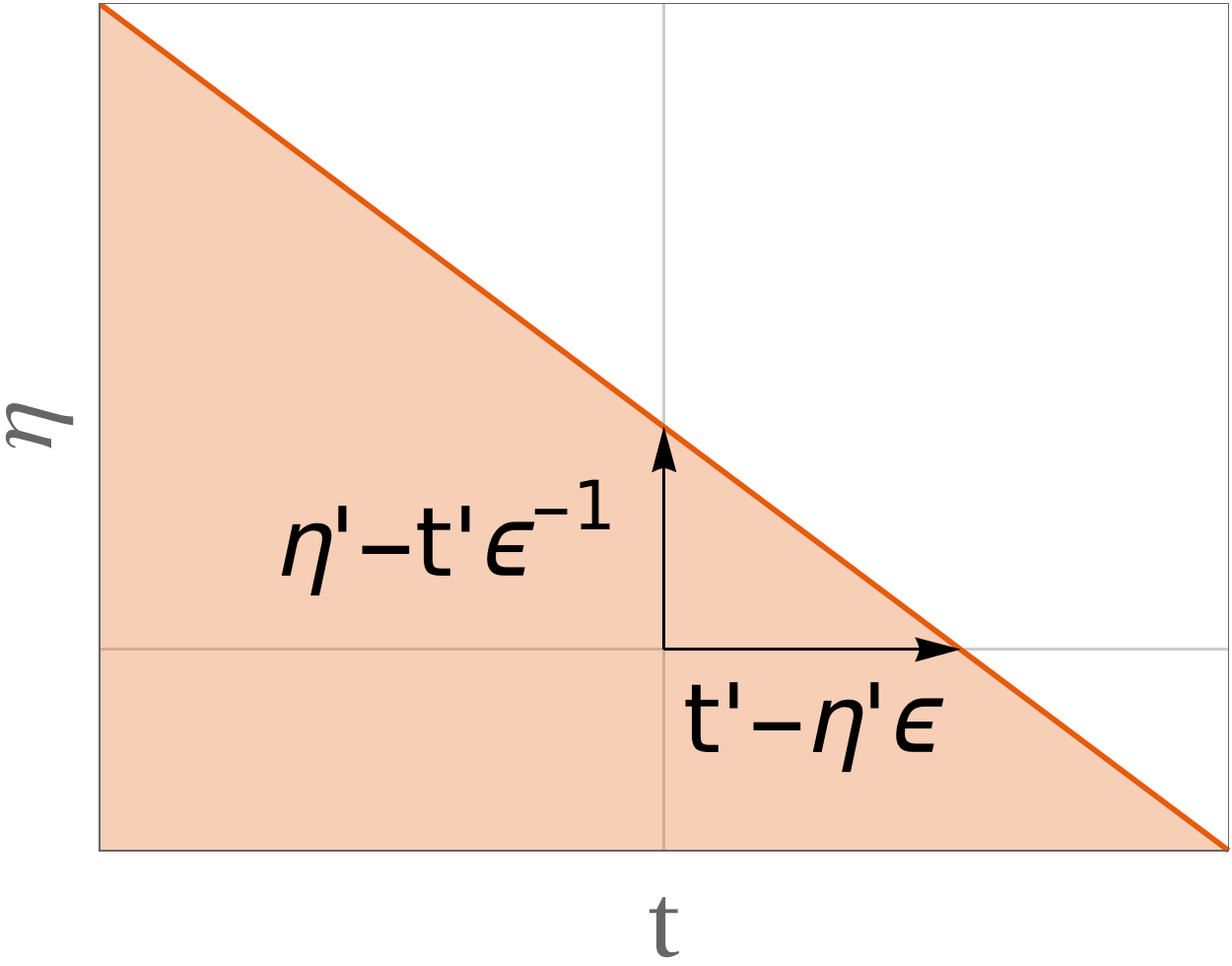}\put(88,71){(b)}
\end{overpic}
\end{minipage}%
\begin{minipage}{.25\linewidth}
\begin{overpic}[width=\linewidth]{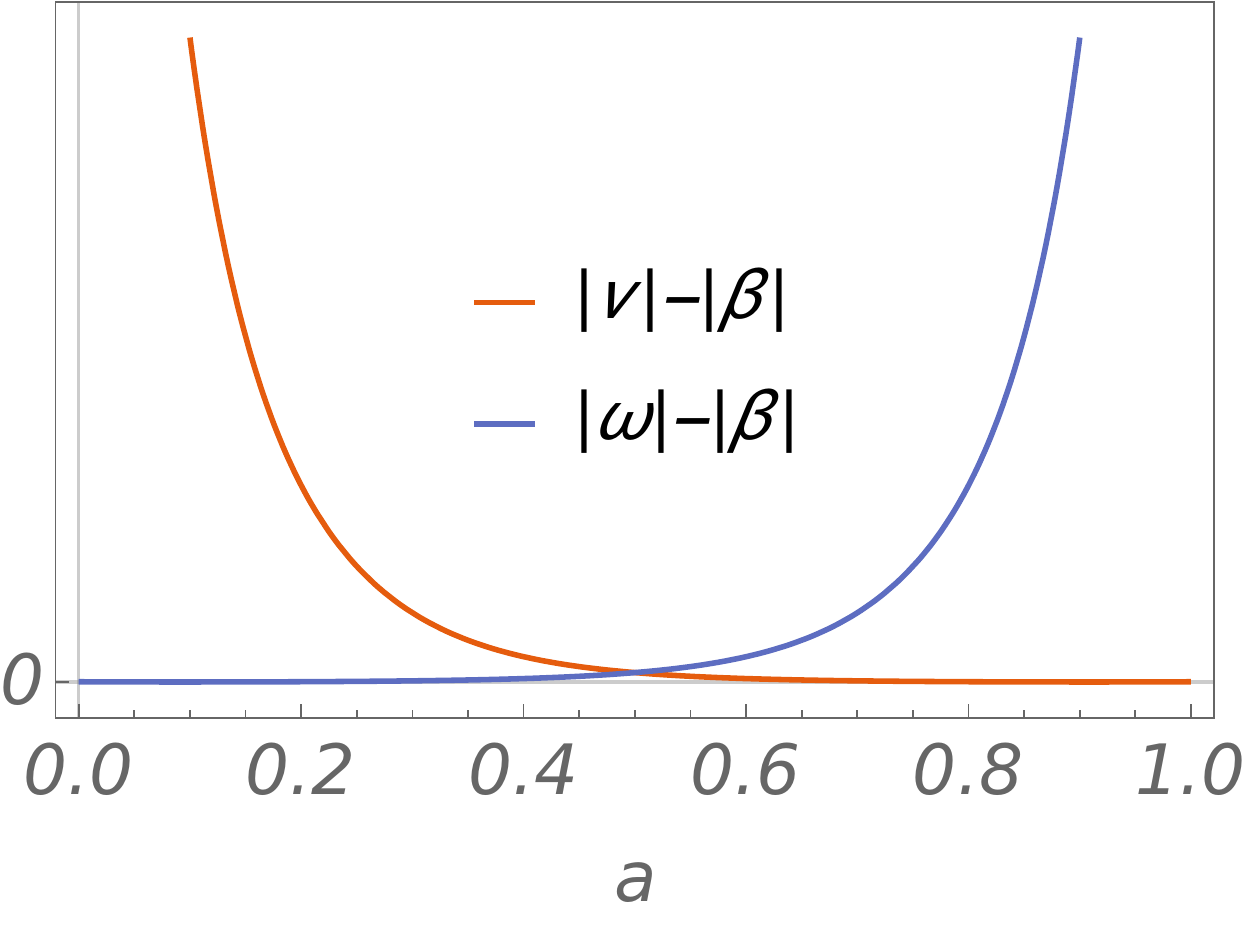}\put(6,61){(c)}
\end{overpic}
\end{minipage}%
\begin{minipage}{.25\linewidth}
\begin{overpic}[width=\linewidth]{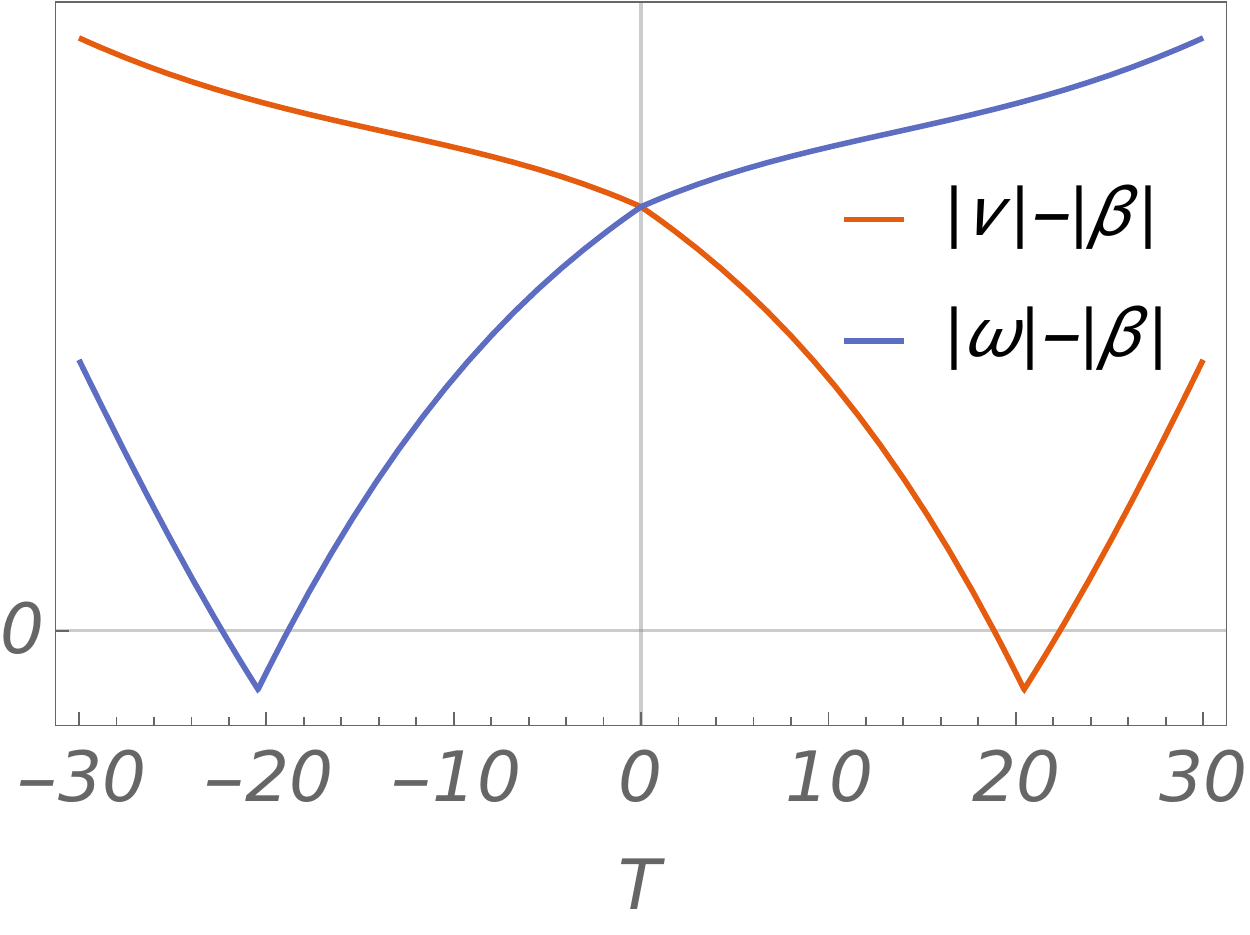}\put(6,61){(d)}
\end{overpic}
\end{minipage}
\caption{(Colour on-line) Panel (a): a density plot of $F(\phi_{\rm min}=\nu^2+\omega^2-\beta^2$ for an arbitrary set of parameters for $\epsilon,t,t',\eta,\eta'$. The black regions, wherein $\nu^2+\omega^2-\beta^2<0$, begin at $\eta/\eta'=\pm1$, $t=\mp\epsilon$. Panel (b): an arbitrary phase diagram of the $\mathbb{Z}$-invariant given by ${\cal W}$ or $\theta_{\cal Z}/\pi$ in the $t-\eta$ plane with $t',\eta',\epsilon$ constant. The shaded and unshaded regions correspond to the non-trivial and trivial regions respectively. Panels (c,d): simultaneous plots of $|\nu|-|\beta|$ and $|\omega|-|\beta$ for the two cases as considered in the main text of: (c) varying the distances between the Dirac-delta potentials as $v=a$, $w=d-a$ with a constant baseline potential $V_0=0$, and (d) varying the baseline potentials are $V_v=-T$ and $V_w=+T$ with $v=w=d/2$.}
\label{fig:phdiag}
\end{figure}

This is true unless it becomes that $|\tau|\gtrsim|\epsilon|$ in which case $|\nu|<|\beta|$ and so $F(\phi_{\rm min})$ becomes negative. This scenario may not occur physically, however, since it would indicate that it is energetically more favourable to execute a hop between lattice sites than it is to remain on the current site. In other words, the basis upon which the tight-binding model is built was poorly chosen. This may be seen most clearly in Fig.~\ref{fig:phdiag}(a) wherein $F(\phi_{\rm min})$ is plotted in the $t-\eta$ plane. The black regions indicate the areas in which $F(\phi_{\rm min})<0$, which begin at the points $t=\pm|\epsilon|$ for $\eta=\eta'=1$ and:
\begin{equation}
t=\frac{\eta(t'\eta'-\epsilon)\pm(t'-\eta'\epsilon)\sqrt{\eta^2+{\eta'}^2-1}}{{\eta'}^2-1},\quad\forall~\eta,\eta'\neq1,
\end{equation}
The other limit is when the bipartition vanishes at the band gap closing point whereat $t=t'$ and $\eta=\eta'$ such that $\nu=\omega$ and $\beta=0$. Clearly, here, $|\nu|>|\beta|$ and $|\omega|>|\beta|$ and so $F(\phi_{\rm min})$ cannot be negative here either.

Moreover, the addition of the further overlap variables, $\eta$ and $\eta'$ within the bulk tight-binding model leads to a modified phase space for the winding number $\mathbb{Z}$-invariant. The band gaps defined through (\ref{eqn:bulkbands}) close at $\nu=\omega$, {\it i.e.} $t-\eta\epsilon=t'-\eta'\epsilon$, and so:
\begin{equation}
\eta=(t-t')\epsilon^{-1}+\eta',
\end{equation}
defines the topological transition point in the phase-space of $t-\eta$. When there is no overlap, in which case $\eta=\eta'=0$, the transition point is $t'=t$, as expected. However, when not so, $\eta$ varies linearly with $t$ with a gradient of $\epsilon^{-1}$ (recall that $\epsilon<0$) and $\eta$-intercept of $\eta'-t'\epsilon^{-1}$ as may be seen in Fig.~\ref{fig:phdiag}(b). This then shines light on one of the important points of this work; namely that the influence of $\epsilon$ is only relevant if there is a finite overlap between neighbouring sites, $\eta,\eta'$. Indeed, within $E(k)$, $\epsilon$ may be ignored as a trivial energy shift if (and only if) $\eta=\eta'=0$. So too here is the topological transition unaffected by $\epsilon$ if $\eta=\eta'=0$.

Within the tight-binding Kronig-Penney model the phase space ought to be restricted to the positive quadrant, $\eta/\eta'>0$ and $t/t'>0$, since there is no effect, akin to a gauge potential or gain and loss, that would alternately change the signs of the nearest-neighbour tight-binding parameters.

As discussed in Section~\ref{sec:bulk}, the argument of the square root in the expression for the bulk energy eigenvalues cannot be negative otherwise the energies becomes complex and exceptional points are introduced to the Brillouin zone. For the two cases as considered and studied within the paper, the behaviours of $|\nu|-|\beta|$ and $|\omega|-|\beta|$ will now be presented to show that it is never that both $|\nu|<|\beta|$ and $|\omega|<|\beta|$ simultaneously.

Recalling that $\nu=t-\eta\epsilon$, $\omega=t'-\eta'\epsilon$ and $\beta=t\eta'-t'\eta$ the relations between $\nu,\omega$ and $\beta$ may be plotted as in Figs.~\ref{fig:phdiag}(c,d) using the tight-binding parameters: $\epsilon=E_0[1+2(e^{-2\kappa v}+e^{-2\kappa w})]$, $t=E_0(3+\kappa v)e^{-\kappa v},\quad t'=E_0(3+\kappa w)e^{-\kappa w}$, $\eta=(1+\kappa v)e^{-\kappa v}$, and $\eta'=(1+\kappa w)e^{-\kappa w}$, for panel (a) and Eqs.~(\ref{eqn:pep}-\ref{eqn:petap}) for panel (b). As may be clearly seen, $|\nu|-|\beta|$ and $|\omega|-|\beta|$ are never both negative at the same point and so $F(\phi_{\rm min})\geq0$ for all range of physical parameters (meaning that $\epsilon,t,t'<0$ and $0\leq\eta,\eta'\leq1$). It is important to note, however, that the fact that $|\nu|-|\beta|$ and $|\omega|-|\beta|$ do become negative in the second case of panel (b) may very well be an indication of the breakdown of the tight-binding approximation there since the bulk bands invert and the finite bands touch when $T\sim\pm(20-25)$. Regardless, however, $|\nu|-|\beta|$ and $|\omega|-|\beta|$ are never simultaneously negative and so $F(\phi_{\rm min})$ is always positive as required.

\end{document}